\documentclass[preprint,authoryear,12pt]{elsarticle}
\pdfoutput=1
\usepackage{amssymb}

\journal{Icarus}

\usepackage{latexsym}
\usepackage{graphicx}

\begin{document}

\begin{frontmatter}

\title{Sunlight refraction in the mesosphere of Venus\\ 
during the transit on June 8$^{th}$, 2004}

\cortext[corresp1]{Corresponding author}
\cortext[deceased]{Deceased on Sept. 11, 2010.}

\author[Cassiopee]{P. Tanga\corref{corresp1}}
\ead{Paolo.Tanga@oca.eu}
\author[LESIA]{T. Widemann}
\author[LESIA]{B. Sicardy}  
\author[WC]{J. M. Pasachoff}
\author[Fizeau]{J. Arnaud\corref{deceased}}
\author[UAI]{L. Comolli}
\author[SAF]{A. Rondi}
\author[SAF]{S. Rondi}
\author[RSAS]{P. S\"utterlin}

\address[Cassiopee]{Laboratoire Cassiop\'ee UMR6202, Universit\'e de Nice Sophia-Antipolis, CNRS, Observatoire de la C\^ote d'Azur, BP 4229, 06304 Nice Cedex 4, France}
\address[LESIA]{LESIA-Observatoire de Paris, CNRS, UPMC Universit\'e Paris 6, Universit\'e Paris-Diderot, 5, place Jules Janssen, 92195 Meudon Cedex, France}
\address[WC]{Williams College Hopkins Observatory, Williamstown, MA 01267-2565, USA}
\address[Fizeau]{Laboratoire Fizeau UMR6525, Universit\'e de Nice Sophia-Antipolis, CNRS, Observatoire de la C\^ote d'Azur, BP 4229, 06304 Nice Cedex 4, France}
\address[UAI]{Gruppo Astronomico Tradatese, Via Mameli 13, 21049 Tradate, Italy}
\address[SAF]{Soci\'et\'e Astronomique de France, 3 rue Beethoven, 75016 Paris, France}
\address[RSAS]{Sterrekundig Instituut, Utrecht University, Postbus 80 000, 3508 TA Utrecht, The Netherlands. Now: Institute for Solar Physics, The Royal Swedish Academy of Sciences, Alba Nova University Center, 106 91 Stockholm, Sweden}

\date{\today}
\begin{abstract}
Many observers in the past gave detailed descriptions of the telescopic aspect of Venus during its extremely rare transits across the Solar disk. In particular, at the ingress and egress, the portion of the planet's disk outside the Solar photosphere has been repeatedly perceived as outlined by a thin, bright arc (''aureole''). Those historical visual observations allowed inferring the existence of Venus' atmosphere, the bright arc being correctly ascribed to the refraction of light by the outer layers of a dense atmosphere. On June 8$^{th}$, 2004, fast photometry based on electronic imaging devices allowed the first quantitative analysis of the phenomenon. Several observers used a variety of acquisition systems to image the event -- ranging from amateur-sized to professional telescopes and cameras -- thus collecting for the first time a large amount of quantitative information on this atmospheric phenomenon. In this paper, after reviewing some elements brought by the historical records, we give a detailed report of the ground based observations of the 2004 transit. Besides confirming the historical descriptions, we perform the first photometric analysis of the aureole using various acquisition systems. The spatially resolved data provide measurements of the aureole flux as a function of the planetocentric latitude along the limb. A new differential refraction model of solar disk through the upper atmosphere allows us to relate the variable photometry to the latitudinal dependency of scale-height with temperature in the South polar region, as well as the latitudinal variation of the cloud-top layer altitude. We compare our measurements to recent analysis of the Venus Express VIRTIS-M, VMC and SPICAV/SOIR thermal field and aerosol distribution. Our results can be used a starting point for new, more optimized experiments during the 2012 transit event. 
\end{abstract}


\end{frontmatter}

\section{Introduction}

Since the mid 18$^{th}$ Century, observers have reported unusual features of the telescopic image of Venus near the inferior conjunction, promptly attributed to its atmosphere. 
Some of them are at the reach of modest instruments (although at small angular distance from the Sun), such as the cusp extension first described by \citet{Schroeter791}, which tends to transform the thin crescent of Venus into a ring of light \citep[e.g.][]{Russell99, Dollfus65}.

One of the most relevant features pertaining to ground-based studies of the Venus atmospheric structure has been observed only during transits, close to the phases of ingress (between 1st and 2nd contact) ot egress (between 3rd and 4th contact) as a bright arc outlining -- 
in part or entirely -- the portion of Venus' disk projected outside the solar photosphere. Traditionally the first account of this phenomenon was attributed to Mikhail V. Lomonosov (1711--1765) who reported his observations of the transit at St. Petersburg Observatory on May 26, 1761~\citep{Marov05}. Actually the poor performance of his small refractor hints that most probably other observers (such as Chappe d'Auteroche, Bergman, and Wargentin) were the first genuine witness on the same date \citep{Link69, Pasachoff12}. 
However, Lomonosov correctly attributed the putative phenomenon to the presence of an atmosphere around the planet, refracting the sunlight in the observer's direction. 

In the following, adopting a denomination widely used in the historical accounts, we will often call this arc ``aureole''. Since both the aureole and the cusp extension occur close to the planet terminator, they are also collectively known as ''twilight phenomena''. For detailed historical reviews the interested reader can refer to \citet{Link69} and \citet{Edson63}. Of course, in this context we neglect the initial motivation for the observation of transits: the determination of the Astronomical Unit by the solar parallax (proposed by E. Halley in 1716), whose interest is purely historical today.

Venus transits are rare, as they occur in pairs 8 years apart, each pair separated by 121.5 or 105.5 years, alternating between descending node (June pairs: 1761/1769, 2004/2012), and ascending node (December pairs: 1631/1639, 1874/1882, 2117/2125). As a consequence, data concerning the aureole are correspondingly sparse and, up to the last event, they have been obtained by simple visual inspection, mainly by using refracting telescopes of modest aperture (typically up to 15-20 cm). This limitation was mainly due to the constraint of organizing complex expeditions including the delicate transportation of all the instruments. 

The 2004 event represents a giant leap in the observation of Venus transits, as the modern imaging technologies available allow for the first time a quantitative analysis of the atmospheric phenomena associated to the transit of Venus. 

We performed measurements of the aureole on the original images obtained through several different instruments, and compared them both with a simple refraction model and with observations obtained in the past. This work summarizes the aspect of Venus close to the Sun's limb during the June 8, 2004 transit as observed by ground--based instruments. A recent paper \citep{Pasachoff11} deals with imaging using NASA's then operating Transition Region and Coronal Explorer solar observatory (TRACE). 

Since the ground-based observations were not specifically organized beforehand, data at our disposal are rather heterogenous. In order to bridge the gap between past visual observations through small telescopes and today's technologies, we decided to consider accounts obtained 
both with professional instruments and low-cost amateur telescopes, either by CCD imaging or
by direct image inspection by experienced observers. In fact, as shown in recent studies of distant solar system objects based on stellar occultation campaigns (e.g. Widemann et al., 2009) while CCD imaging offers today the most valuable quantitative measurements, small telescopes and visual observations allow a significant increase of constrains on the phenomena  - and in the case of Venus, the most direct comparison to past reports. The results obtained from the analysis of the most significant image sets, representing a certain variety of instrument size and quality, are illustrated in this paper.

The paper is organized as follows. First, we describe the conditions of the 2004 event, and the reconstruction of limb geometry (section~\ref{S:geom}). We then provide an extensive review of the measured spatial and temporal variations for the brightness of the aureole in wide or narrow-band photometry (section~\ref{S:obs} and \ref{S:color}). We then address the basic physical principles of the atmospheric differential refraction model producing the aureole (section~\ref{S:model}) and we use it for the interpretation of the observations (\ref{S:application}). In Section \ref{S:VeX}, the modeling of the aureole is compared to recent analysis of the Venus Express observations (VIRTIS-M, VeRa and SPICAV/SOIR) regarding the thermal field and cloud-deck altitude and haze distribution pertaining to this study, as well as ground-based mid-infrared spectroscopy of non-LTE CO$_2$ emission. The comparison is discussed.

\section{Geometry of the transit in 2004}
\label{S:geom}
For simplicity and following \citet{Link69} hereinafter we call ``phase'' 
($f$) of the event the fraction of Venus diameter external to the solar photosphere. 
A value $f$=0 corresponds to the planet entirely projected on the Sun, tangent 
to its limb. When $f$=0.5, the planet centre will be exactly on the solar limb, and so on.

\begin{figure}[hbp]
\centerline{\includegraphics[width=0.95\textwidth]{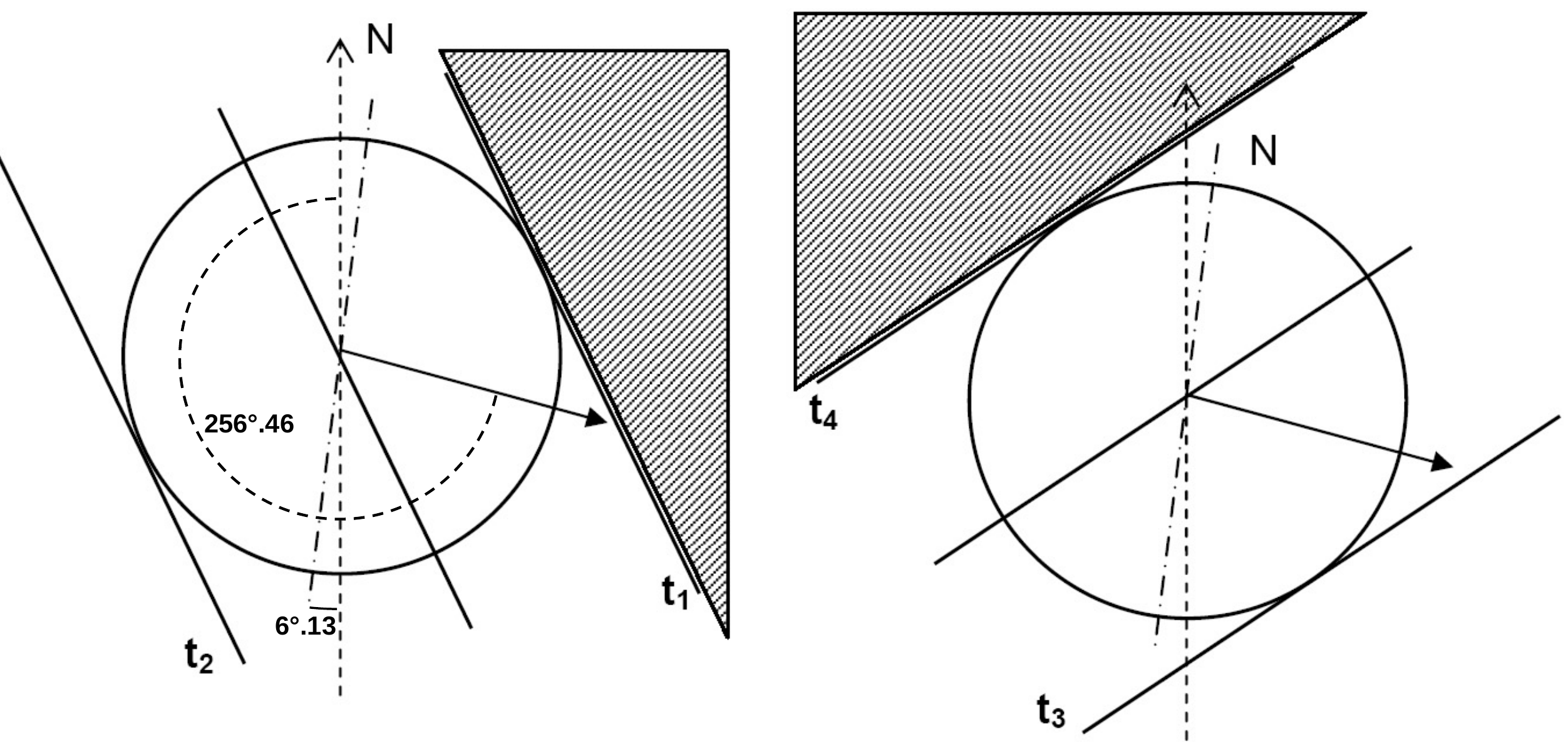}}
\caption{Sketch of Venus' disk orientation at the ingress (left panel) or egress (right) of the 
transit. The greyed area corresponds to the Sun's photosphere at the epoch when the Venus disk is 
externally tangent to the Sun (first and fourth contact: t$_1$ and t$_4$ respectively). The Solar limb
is also indicated at the second and third contact (labeled t$_2$ and t$_3$) and when projected on the center of Venus. The solid arrows indicate the direction of the apparent motion of the planet relative to the Sun. The vertical 
dashed line corresponds to the sky North-South direction, while the dash-dotted one represents the 
sky-projected rotation axis of Venus.}
\label{F:orientation}
\end{figure}

In Fig.~\ref{F:orientation} the orientation of the disk of Venus relatively to the solar limb is given, both for ingress and egress phases. In both cases 
the temperate latitudes are tangent to the Sun for $f=0$ and $f=1$. At first 
contact the South pole of Venus remains projected longer on the sky, while 
the North pole is the first one to enter on the solar disk\footnote{In the following we will always 
use the IAU convention i.e. the North pole is the one lying on the northern side 
of the ecliptic.}.
The sequence is inverted between third and fourth contact, such that it is always 
the South pole to be observed externally to the Solar disk for a longer time. 

The total limb crossing for the disk of Venus lasted 18.9 minutes and the apparent radius of the planet was 28.9 arcsec.

\section{Observations and measurements}
\label{S:obs}

\begin{figure}[hbp]
\centerline{\includegraphics[width=0.95\textwidth]{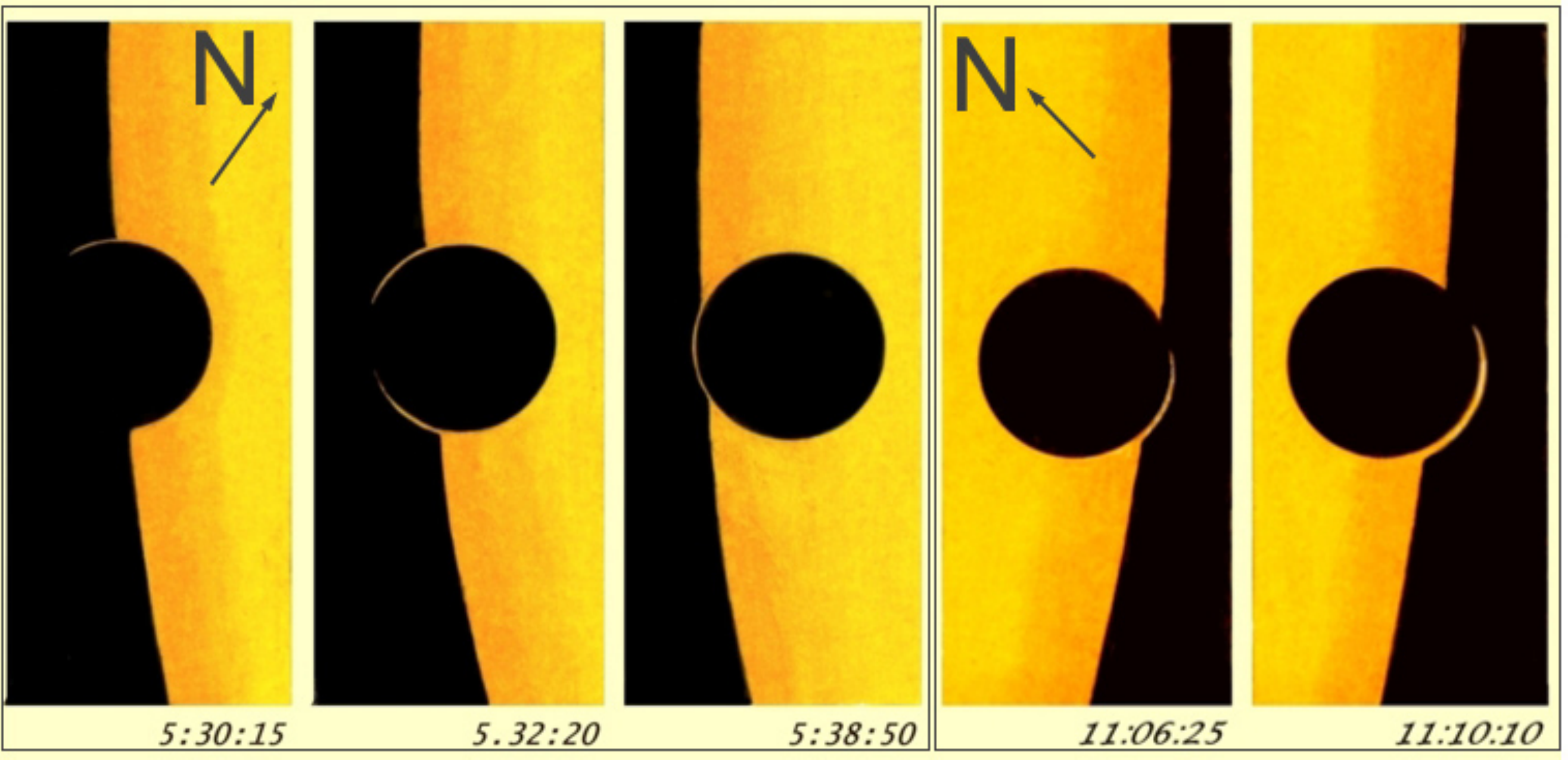}}
\caption{Drawings of Venus at the beginning (three leftmost panels) and end of the event (remaining two panels), as seen visually by an expert amateur astromer (Mario Frassati, Italy) through a 20~cm Schmidt-Cassegrain telescope. Courtesy of the archive of Sezione Pianeti, Unione Astrofili Italiani.}
\label{F:frassati}
\end{figure}

The European observers providing
the data sets described further on had particularly favourable conditions around the 
end of the event, while the ingress of the planet on the solar disk was observed at much higher airmass, i.e. at low elevation above the horizon. However, visual observers under good sky conditions and employing a magnification higher than $\sim$150$\times$ had no particular difficulty in identifying the bright aureole outlining the Venus disk between 1st and 2nd contact, while it was crossing the Solar limb (i.e. for f$<$1). Skilled observers immediately noticed the non-uniform brightness of the aureole along the planet disk (a high-quality drawing by an expert amateur observer is shown in Fig.~\ref{F:frassati}).

For a quantitative analysis of this phenomenon we rely on images obtained
by CCD cameras through different telescopes. Given the casual nature of most images of the aureole that have been produced during the event through instruments of all sizes, not all of them are suited for a comprehensive analysis. We thus selected the representative sample of observations presented in the following, with appropriate image quality and information content. Table~\ref{T:obs} summarizes the actual contributions that were collected and analyzed, as presented in the following. 

\begin{table}[t]\scriptsize
\caption{Measured observations. The availability of ingress ("In" column) or egress ("Ex") images is given. The number of measured images follows. These are {\sl final} images, as a result of an averaging process when needed. The final column identifies the filter name and/or their central wavelength/width. See case by case detailed explanations in the text. \label{T:obs}}
\begin{tabular}[t]{lllcccc}
\hline
{\bf Observer} & {\bf Instrument} & {\bf Site} & {\bf In} & {\bf Ex} & {\bf N. images} & {\bf $\lambda$/FWHM (nm)}\\
\hline
J. Arnaud & Themis & Tenerife (Spain) & - & X & 50 & V\\
L. Comolli & 20 cm Schmidt-Cass. & Tradate, Italy & X & X & 50 & V\\
P. Suetterlin & DOT & La Palma (Spain) & & X & 21 & 430.5/1.0, G\\
& & & & X & 21 & 431.9/0.6, Blue cont.\\
& & & & X & 21 & 655.0/0.6, Red cont.\\
& & & & X & 21 & 396.8/0.1, Ca II H\\
S. Rondi & 50 cm Tourelle refr. & Pic du Midi (France) & X & & 32 & NaD1/10 \\
\hline
\end{tabular}
\end{table}

\begin{figure}[hbp]
\centerline{\includegraphics[width=0.9\textwidth]{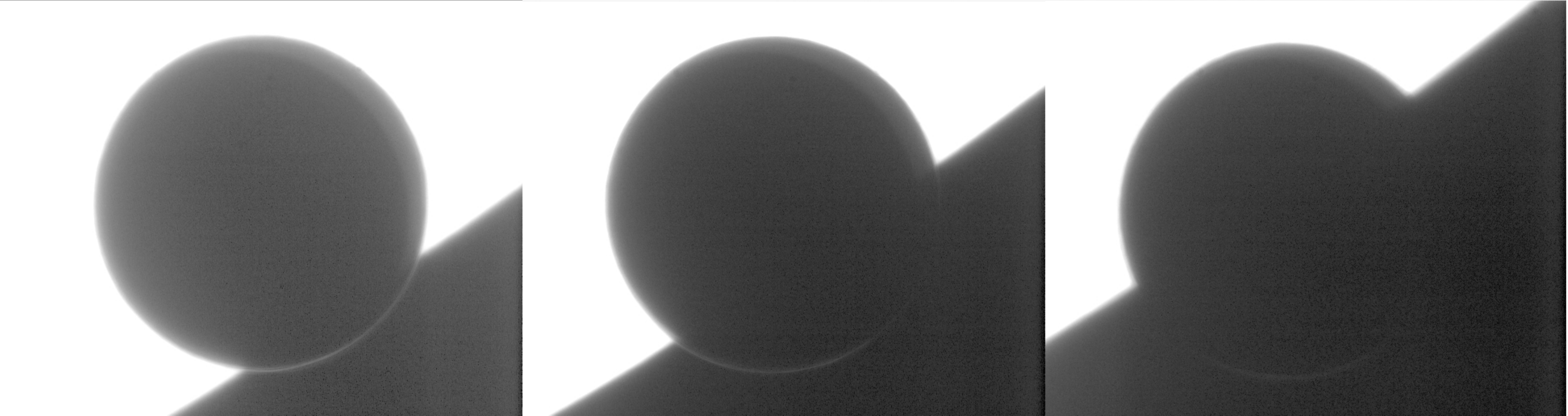}}
\caption{Three images concerning the final phases of the transit obtained by the DOT telescope in the ''G'' band. Each image has been altered in contrast (gamma = 0.4) in order to better show the thin and faint arc of light outlining (partially or completely) the Venus limb. The orientation is the same as in Fig.~\ref{F:orientation}. The bright saturated area is the solar photosphere. The progressive reduction in extension and brightness of the aureole with time is clearly visible.}
\label{F:DOT_seq}
\end{figure}

Images have undergone a standard calibration process (dark current and flat--field correction).  The brightness of the aureole, when present, was measured. Fig.~\ref{F:DOT_seq} shows an example of CCD image sequence between 3rd and 4th contact on June 8, 2004. The bright arc or aureole is clearly visible in the three frames obtained at an increasing distance of the Venus disk center from the Sun. For illustration purposes a contrast enhancement is applied. The inhomogeneity of the aureole is clearly visible on the motionward limb of Venus.

\begin{figure}[hbp]
\centerline{\includegraphics[width=0.7\textwidth]{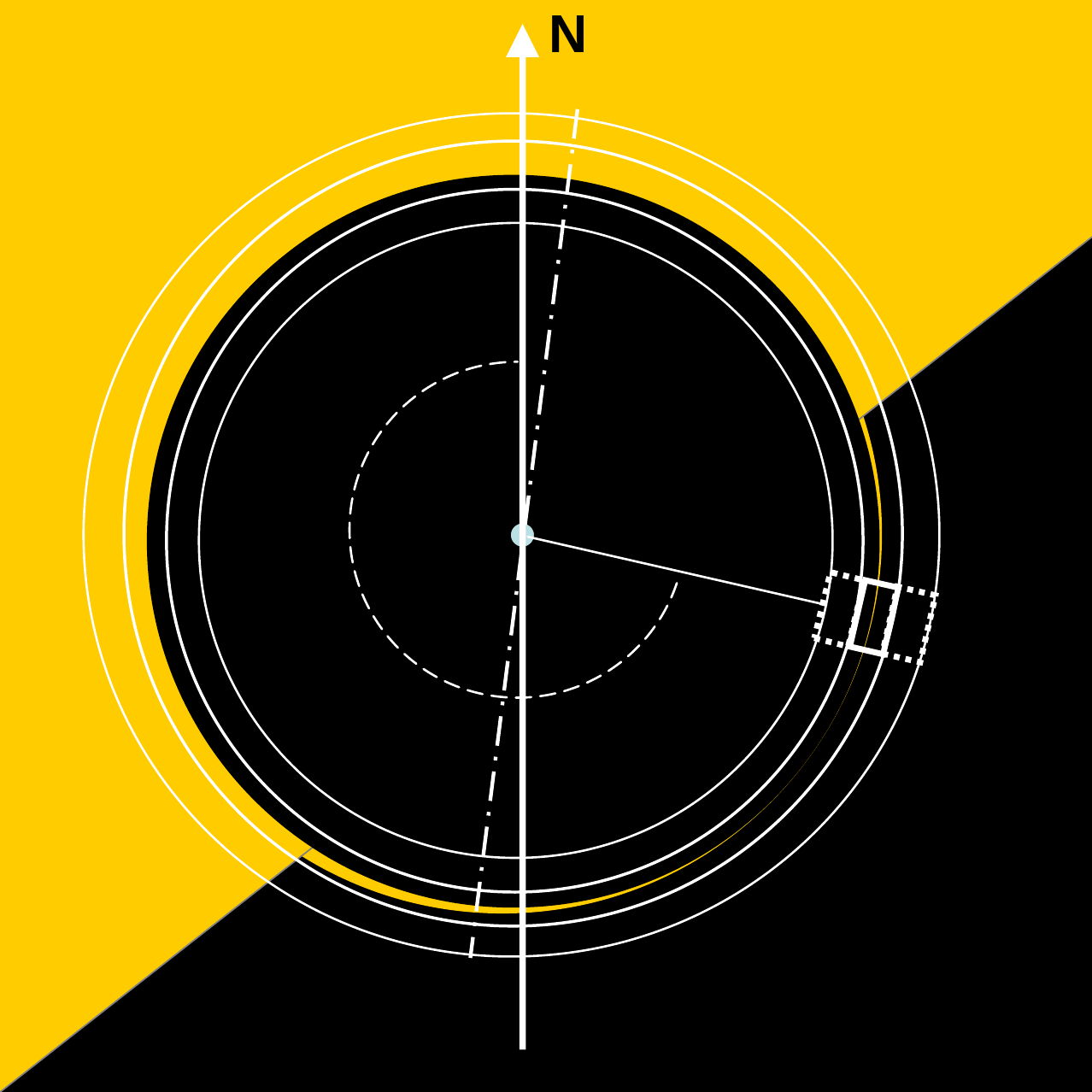}}
\caption{Scheme of the procedure adopted for the measurements and implemented in a custom software. Here black is used to represent the sky background and the disk of Venus. The four white circles define three concentric rings, with the aureole entirely contained in the central one. The nearby external and internal rings are used for evaluating the background. Each ring is divided in sectors of equal angular size. One of them is represented for the aureole (solid rectangle) and the corresponding background areas (dotted). The dashed angle represents the position angle (relative to the celestial North) used for representing the brightness profiles in following plots. The dashed-dotted line represents the orientation of the rotation axis of Venus.}
\label{F:pdm}
\end{figure}

The arc's photometric profile was obtained by the integration of the signal in a ring centered 
on the planet, outlined in Fig.~\ref{F:pdm}. The full ring is divided into sectors of identical angular extension (as seen from the center of the Venus disk), 
each corresponding to a flux integration area. Among them, only those sectors projected on the sky are then considered.
The background is evaluated on two rings, one on the inside and the other on 
the outside of the main ring. Their outer and inner edges (respectively) are also plotted in Fig.~\ref{F:pdm}.
Each background ring is divided in sectors as the measurement annulus. 
This way, each patch on the main annulus is accompanied by two background patches at the same position angle (relative to the Venus disk center) but on opposite sides. Their average flux, weighted by their surfaces, yields the final background value to be used for the given measurement patch. 

In the images we selected, the solar disk is not saturated, so a normalization relative 
to a selected photospheric region is possible. We decided to use the value of brightness measured at 1 Venus radius from the solar limb. On the images, the geometric limb is estimated by the analysis of the radial brightness profile of the solar disk. 
The photometric normalization value is provided by the average of the flux in 
a square window of the same surface as the sectors used for measuring the aureole brightness. A final normalization is done for deriving the plotted values, corresponding to the flux coming from an unresolved, thin segment of aureole 1~arcsec in length, normalized to the brightness of a square patch of photosphere having a surface of 1~arcsec$^2$.

In the following, we present technical details of each data set, with the corresponding aureole luminosity derived. The error bars given in the plots are computed by considering the Poissonian statistics of the photometric noise. We indicate by $N_s$ and $N_b$ the
number of photoelectrons coming from the aureole and the background,
respectively, in each arc bin. The uncertainty that we consider is $\sigma = (N_s+N_b)^{0.5}$. In our case, $N_b \gg N_s$, thus the only relevant contribution to noise comes from the background, and is a function of the
distance of the arc element from the solar limb. The camera read-out-noise is cleary smaller than the background contribution and can be discarded as well. The same applies to noise associated to the dark current, since exposure times are extremely short.

From a practical point of view, we have verified that $\sigma$ appears to be fairly
constant over the entire arc, growing of about 50$\%$ only for measurement
points very close to the solar limb. For this reason, we choose to show only one error bar, common to all curves, except in the case of observations obtained at the DOT telescope which are used for modelling the aureole flux. In the related plots, two error bars separately for low and high $f$ values, provide an additional information on the amount of noise variation in time.



\subsection{Tourelle Solar Telescope - Pic du Midi, France}

Only the ingress has been imaged by this station, since clouds have prevented a clear view of the final phases. 
Given the image scale, the radius of Venus is 142.5 pixels.
For a set of selected images in which the aureole is well visible at the visual 
inspection, its integrated flux was determined using different widths of
the measuring ring, from 2 to 14 pixel. The resulting "growth curve" \citep{Howell89}
shows the convergence toward the inclusion of the complete flux coming from
the arc. The flux was considered to be completely inside the ring at a width of
9 pixel, then used for all measurements. The arc was sampled at angular steps of 9 degrees. 

Also, it was found that due to the faintness of the aureole, a useful improvement of the SNR was obtained by averaging images by groups of 3. The final image set obtained this way was much easier to measure than the
original single frames. In the following discussion of Pic du Midi data we call ''images'' the members of this final set.

\begin{figure}[htbp]
\centerline{\includegraphics[width=0.95\textwidth]{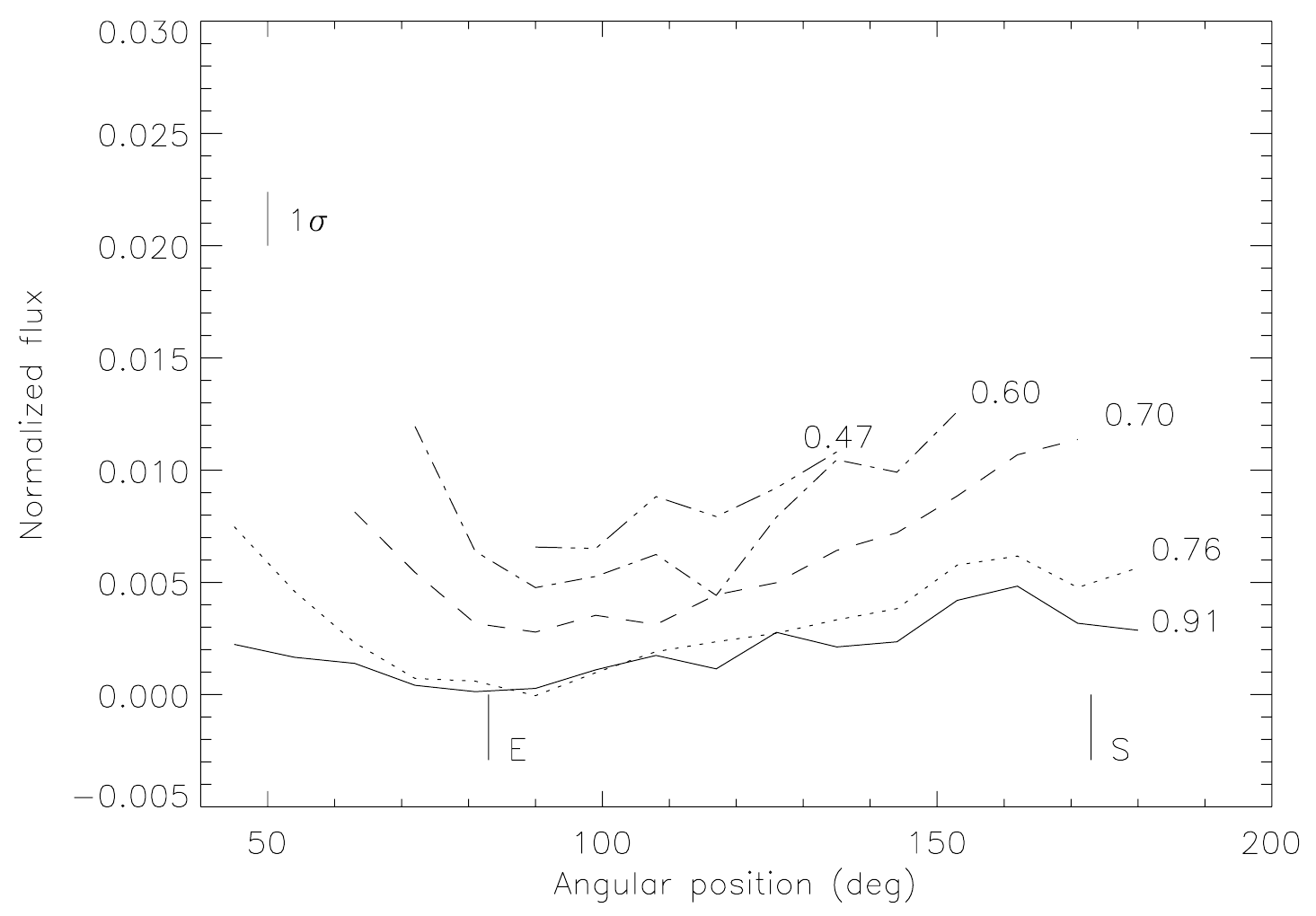}}
\caption{Photometric profile of the aureole during ingress, obtained at the Tourelle telescope, Pic du Midi. 
The plotted value represents the flux coming from a segment of 1~arcsec of the
aureole, normalized to the flux coming from a 1~arcsec$^2$ surface of the solar photosphere at 1 Venus radii from the solar limb (measured on the same set of images). The horizontal axis shows the position angle along the Venus limb, with the usual conventions in the equatorial reference: origin of angles at the celestial N, increasing toward E.  
The two vertical lines represent the position angle of the equator (left, ''E'' label) and
the South pole (right, ''S''). 
The flux from the aureole increases with time. Upper curves, with the planet about halfway across the solar limb, cover a short angular range since their extremes are strongly contaminated by a high background. Labels indicate the values of $f$ associated to each curve.
}
\label{F:tourelle1}
\end{figure}

The arc is present on 16 images, and a photometric profile was derived for each. 
Given the poor signal-to-noise ratio (SNR) the curves were later average by four, obtaining the result 
presented in Fig.~\ref{F:tourelle1}. The profiles have been trimmed in order to
exclude the region in which the measurement areas (main or background patch) are
contaminated by the photosphere. In other words, only the portion corresponding to
the arc projected on the sky is plotted. In the case of the Tourelle telescope, the background strongly contaminates the signal when the planet is about halfway into the ingress. We thus had to be very conservative concerning the fraction of arc to be considered.

Despite the averaging, the result is still affected by a significant noise, as indicated by the size of the vertical bars in Fig.~\ref{F:tourelle1}, representing the 1-sigma level. 
For this reason, small scale fluctuations in the brightness profile probably do not 
have a strong statistic significance if considered separately. However, some general trends are present and repeat from one curve to the other. For example, a general slope of the curve bottom is present. Also,
at large values of $f$ (lowest curves) a maximum in intensity is detectable 
close to the South pole of the planet. Interestingly, the maximum is not exactly centered on the pole, a 
feature that repeats in the other data sets described further on. 
The intensity minimum coincides rather well with the position angle of the 
equator of Venus (bar on the left). Even at later times, represented by the curves at 
the highest values, the trend of growing brightness toward the pole is present. 
In all cases, if we neglect the extreme tips of the arc very close to the photosphere,
the aureole flux is less than about 2\% the photospheric value, and is detectable
down to about 0.2\%.

\subsection{Tradate (20 cm Schmidt Cassegrain) data}

\begin{figure}[htbp]
\centerline{\includegraphics[width=0.95\textwidth]{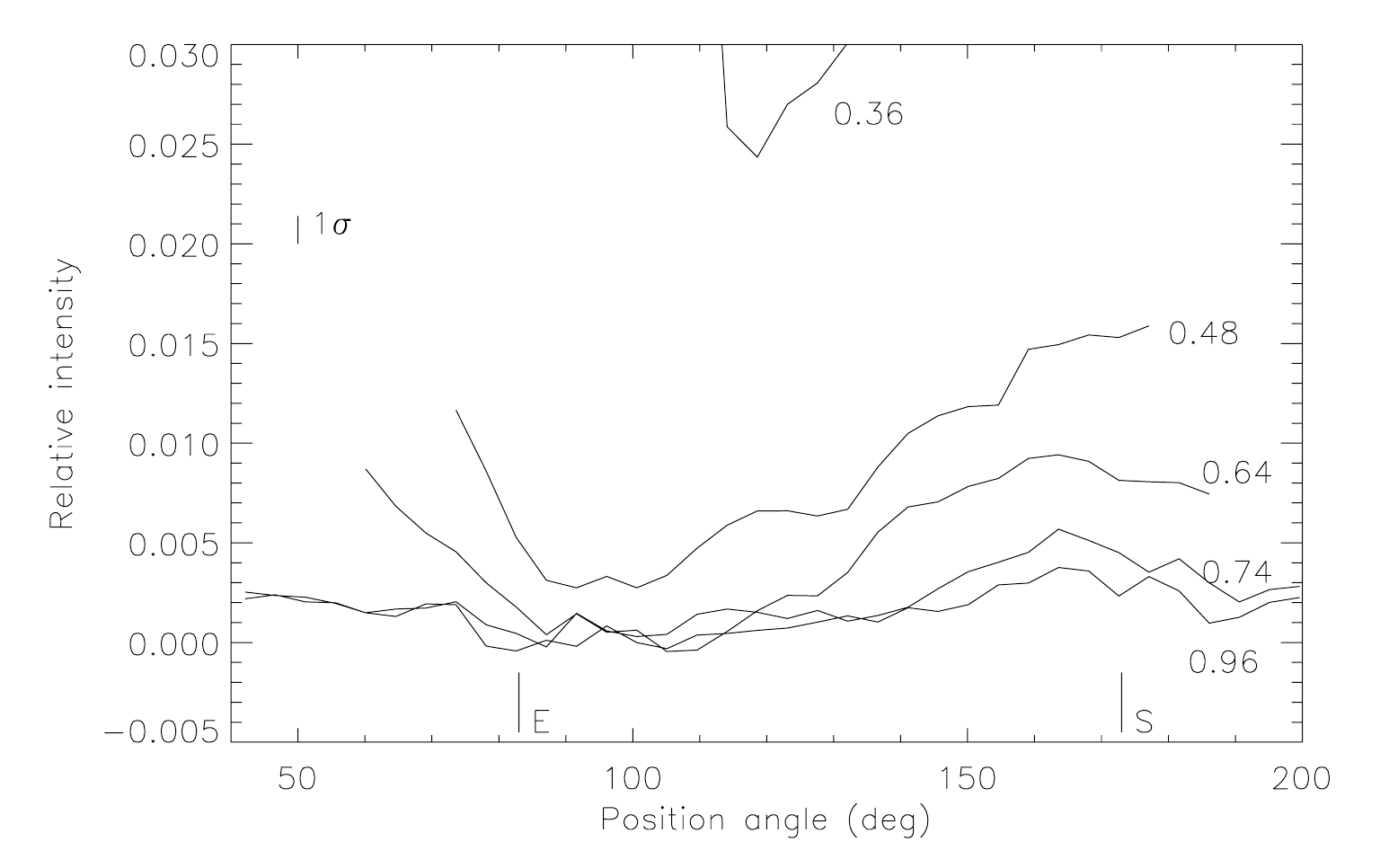}}
\caption{Photometric profile of the aureole during ingress, obtained by
L. Comolli, normalized as the other image sets. Vertical bars indicate the
entity of the 1-sigma uncertainty for each data point. The left bar represents
the position angle of the equator, while the right bar refers to the South
pole. Labels indicate the values of $f$ associated to each curve.
}
\label{F:comolli_i}
\end{figure}

L. Comolli observed the event by a digital video-camera, with exposure times
variable between 1/2000 and 1/50 sec. Only the shortest exposures have been
used, since they do not present saturation on the solar photosphere, thus
allowing a normalization equivalent to those presented in the other cases. 
Each analyzed image is the result of the sum of the best 200-300 frames over
$\sim$1~minute of recording.

Both ingress and egress were imaged and analyzed, using a step of 4.5 degrees
over the Venus limb. The measured flux distribution during egress is well
consistent with DOT data (presented below), but limited by a far lower resolution. 
For this reason they are not plotted here.
More interesting is the comparison of the ingress luminosity profiles to the results 
obtained by the Tourelle Solar Telescope images.

Five profiles were obtained and are presented in Fig.~\ref{F:comolli_i}. They
confirm, from both the point of view of the distribution and the flux values,
the data obtained at the Tourelle telescope . Discrepancies around 15$\%$ at f$\sim$0.5 in the 
faintest part of the aureole appear small when taking into account the wide difference
in equipment and site,
however they clearly hint to the difficulty of obtaining clean and 
appropriate signals at such small solar elongations.

\subsection{Themis Solar Telescope - Pico del Teide Observatory, Tenerife, Spain}

Themis data are complementary to the Pic du Midi observations, since from this site 
it was the exit phase to be imaged, both through a narrow-band H-alfa filter 
and a filter centered at 430 nm.
The image sampling is lower (Venus radius~$\sim$~70 pixels) but a large 
image rate is available (1 image each 1.06 seconds). The sequence was thus divided 
into different sets of images, containing 20 to 60 frames each. The most populated 
sets correspond to the largest values of $f$, containing an extremely faint 
aureole slowly fading into the background.

Inside each set, each image was re-sampled with a re-alignment 
of the disk of Venus relatively to an image chosen as reference. 
An automated procedure based on the computation of cross-correlations by Fast Fourier Transforms 
was used, providing the shift to apply to each image. Small relative image drifts are thus minimized and the SNR of 
the faint arc is preserved during the summing procedure that yields a final image for each set. 

\begin{figure}[htbp]
\centerline{\includegraphics[width=0.7\textwidth]{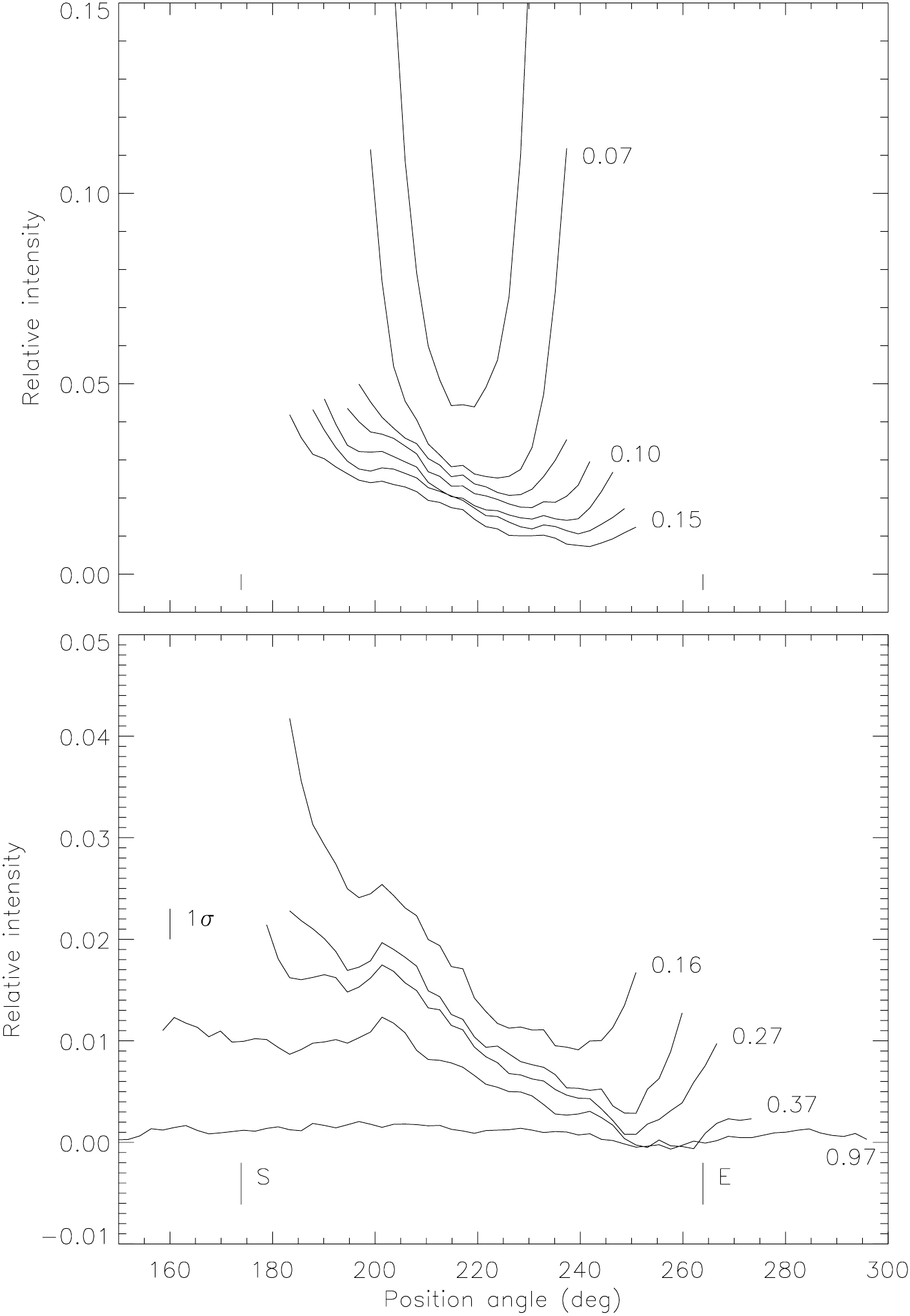}}
\caption{Photometric profile of the aureole during egress, obtained at the Themis solar telescope, Tenerife. 
The intensity is normalized as in Fig.~\ref{F:tourelle1}.
The two vertical lines represent the position angle of the equator (right) and
the South pole (left). The curves have been split into two families: the upper panel 
refers to the beginning of the egress, while the bottom one, with expanded
vertical scale, refers to the central and
final phases. The labels indicate the values of $f$ associated to the corresponding curves.}
\label{F:themis1}
\end{figure}

In this way, 50 images were derived from the sum of the frames 
inside each set, and measured as those of the previous section. 
The normalization procedure adopted was also the same. Due to 
poor seeing, a width of 7 pixels 
was needed to include the entire flux. On the other hand, the good 
SNR allowed to use a step of 2.25 degrees (corresponding to 
2.75 pixels), in order to avoid losing information about small 
scale variations. However, the inspection of the resulting curves did
not reveal any clear feature at that resolution 
level, much smaller than the arc width smearing due to 
turbulence.
A smoothing by running box on three points was thus applied, and 
the final curves 
are presented in Fig.~\ref{F:themis1}. 
The overall brightness of the aureole results to be 5\% of the photospheric reference at maximum, with the shortest measured arc. 
The results at Themis shows some features that are symmetric to those observed 
at the Tourelle telescope during the entry. Here, the brightness increases from the equator (right bar in the plot, label ''E'') toward the pole 
(left bar, ''S''). Again, when the polar region is observable projected on 
the sky, a maximum in brightness is detected close to the pole, at 
about 25 degrees from it. This is the last portion of the aureole 
that remains visible.
Another interesting feature is the relative maximum at position angle $\sim$200$^\circ$, 
(latitude $\sim$-65$^\circ$) which is constantly present during most of the evolution of the arc profile. 

\subsection{DOT telescope - Roque de Los Muchachos observatory, La Palma, Spain}
\label{S:DOT}

\begin{figure}[htbp]
\centerline{\includegraphics[width=0.7\textwidth]{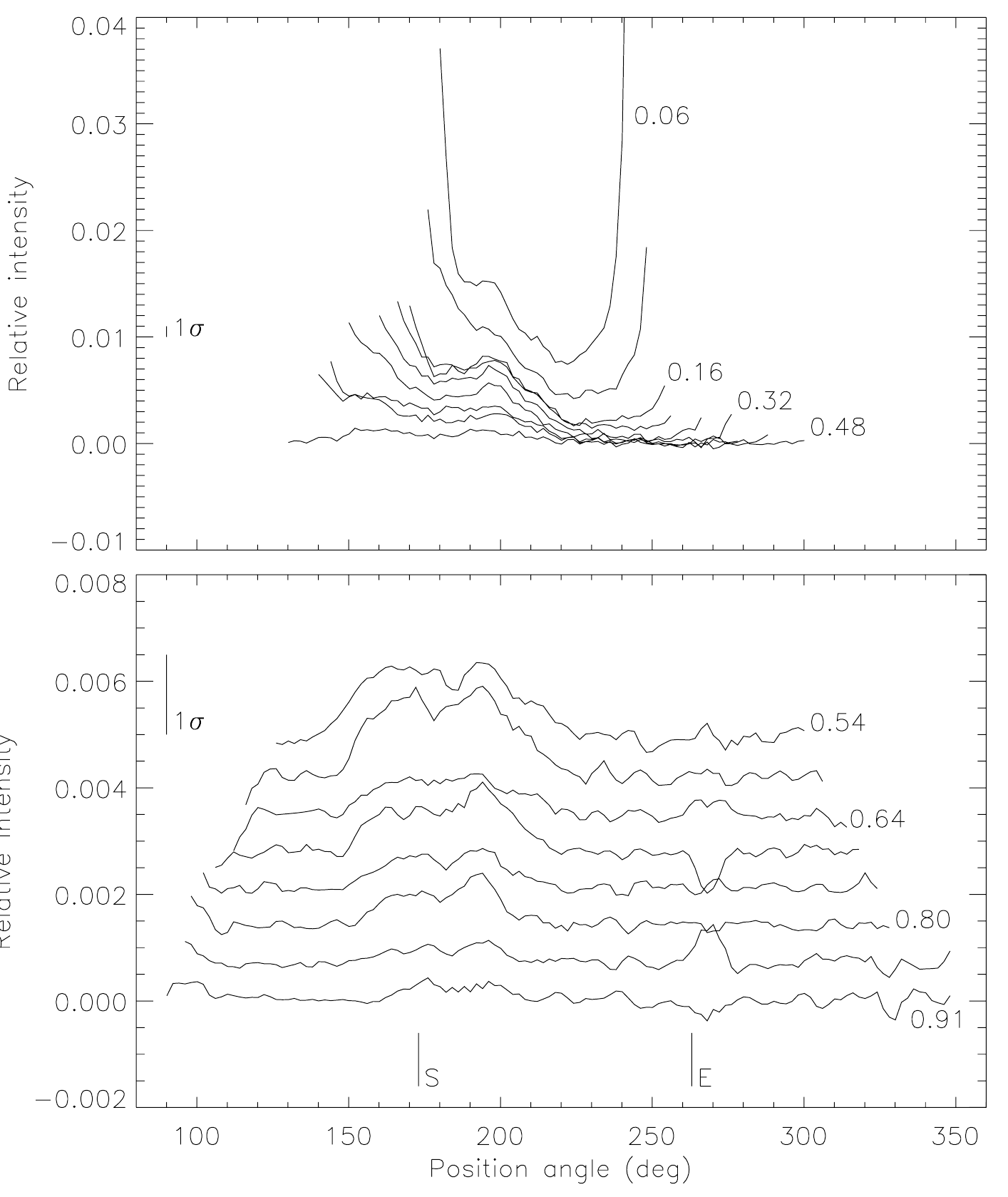}}
\caption{Photometric profile of the aureole during egress, obtained at the
DOT solar telescope, La Palma, in the G band. The intensity is normalized as in Fig.~\ref{F:tourelle1}.
The two vertical lines represent the position angle of the equator (right) and
the South pole (left). The curves have been split into two families: the upper panel 
refers to the beginning of the egress, while the bottom one, with expanded
vertical scale, refers to the central and final phases. The time interval between
each curve is 60 seconds. For an easier visibility of the bottom
panel, each curve has a step of 7$\times$10$^{-4}$ added in intensity relatively to
its lower neighbor. The labels indicate the values of $f$ associated to the corresponding curves.
Two levels of uncertainty ($\sigma$) related to a position angle in the range 150-250$^\circ$ are given in the two panels. The upper one is applicable to values of $f<$0.4, while the other one refers to $f$=0.7. }
\label{F:DOT_Gband}
\end{figure}
\begin{figure}[htbp]
\centerline{\includegraphics[width=0.7\textwidth]{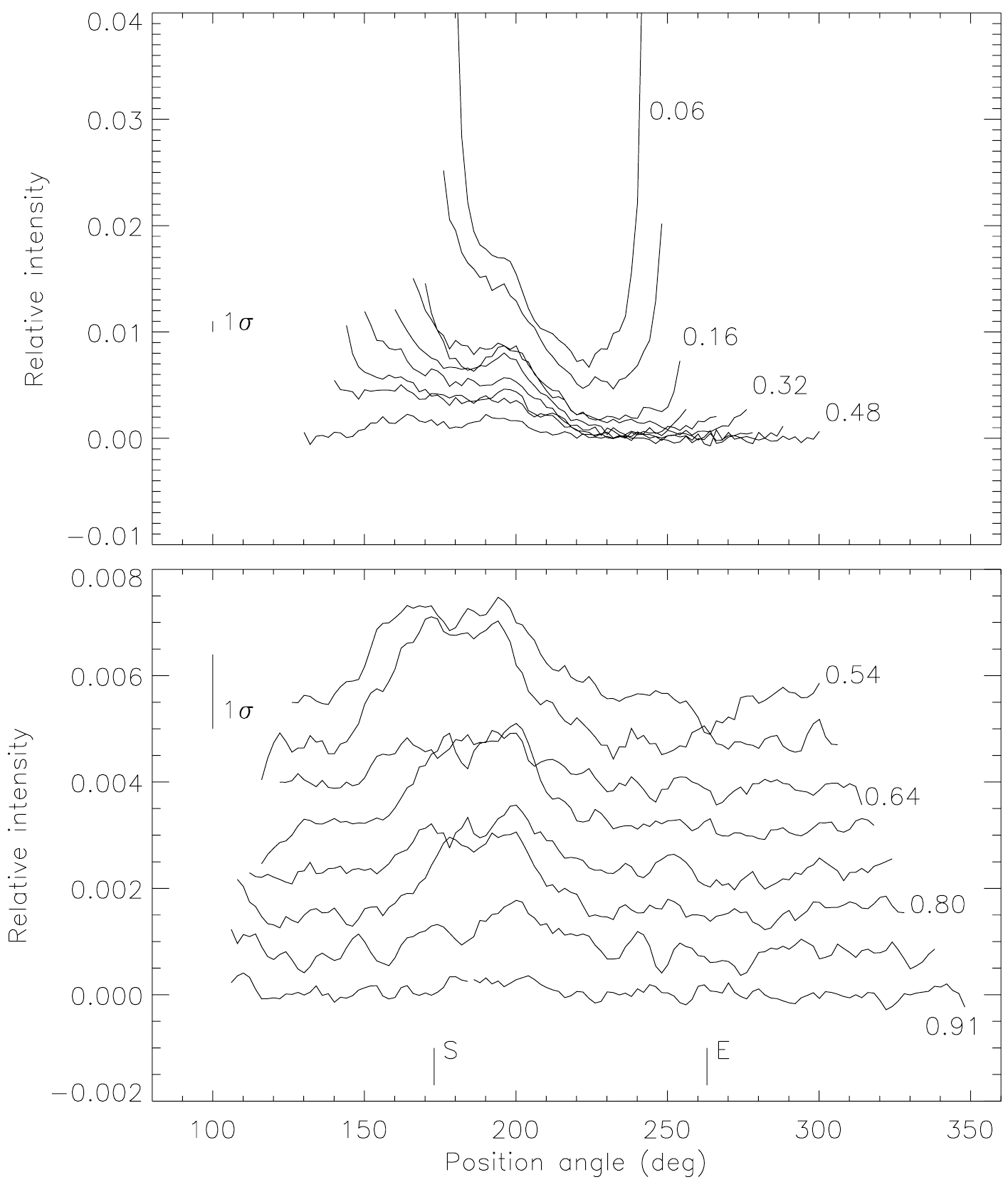}}
\caption{As in Fig.~\ref{F:DOT_Gband} in the blue continuum at 431.9 nm.}
\label{F:DOT_BlueCont}
\end{figure}
\begin{figure}[htbp]
\centerline{\includegraphics[width=0.7\textwidth]{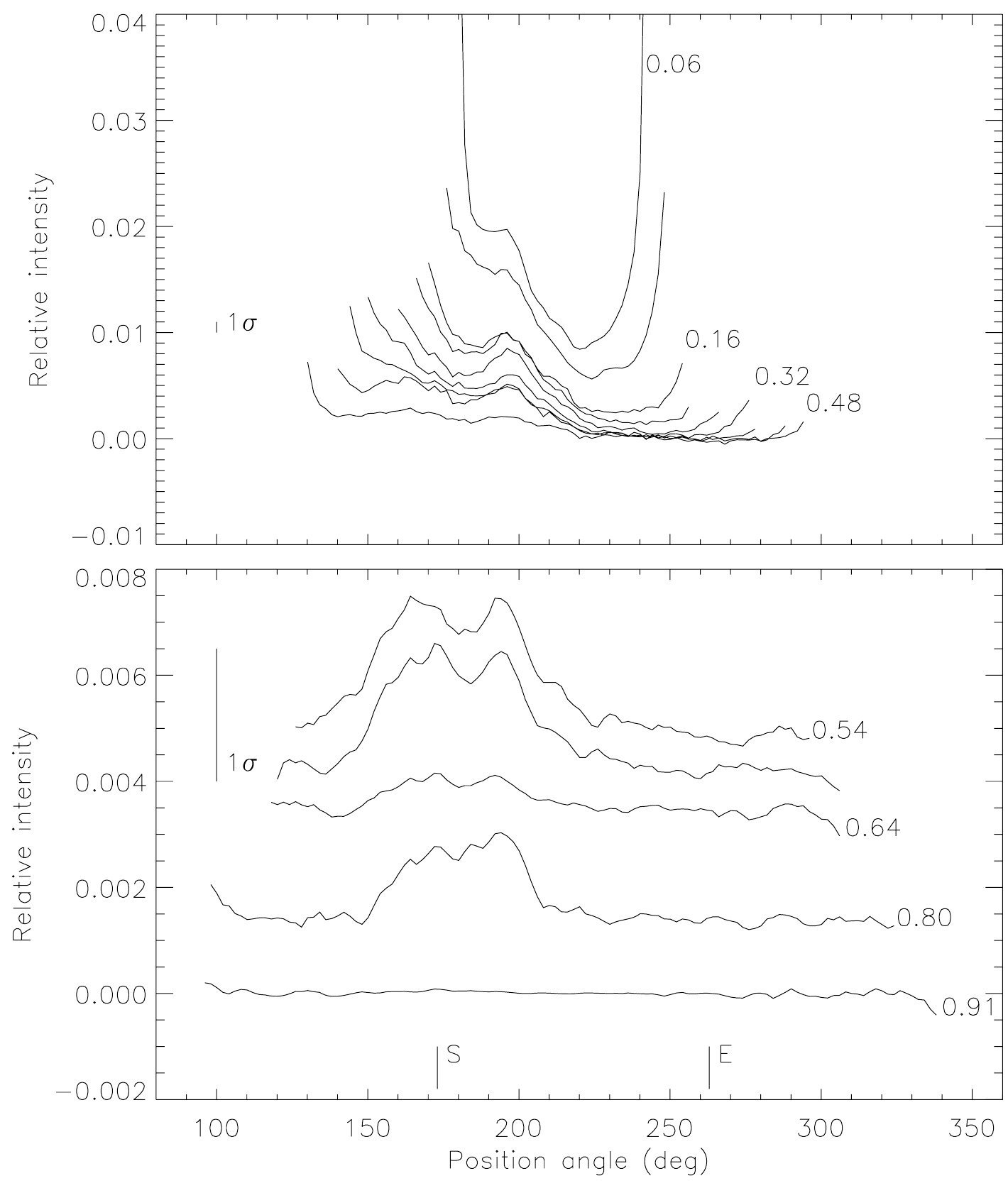}}
\caption{As in Fig.~\ref{F:DOT_Gband} in the red continuum at 655 nm. Relative to the other wavelength bands, here 3 curves are missing due to corrupted data files.}
\label{F:DOT_RedCont}
\end{figure}
\begin{figure}[htbp]
\centerline{\includegraphics[width=0.7\textwidth]{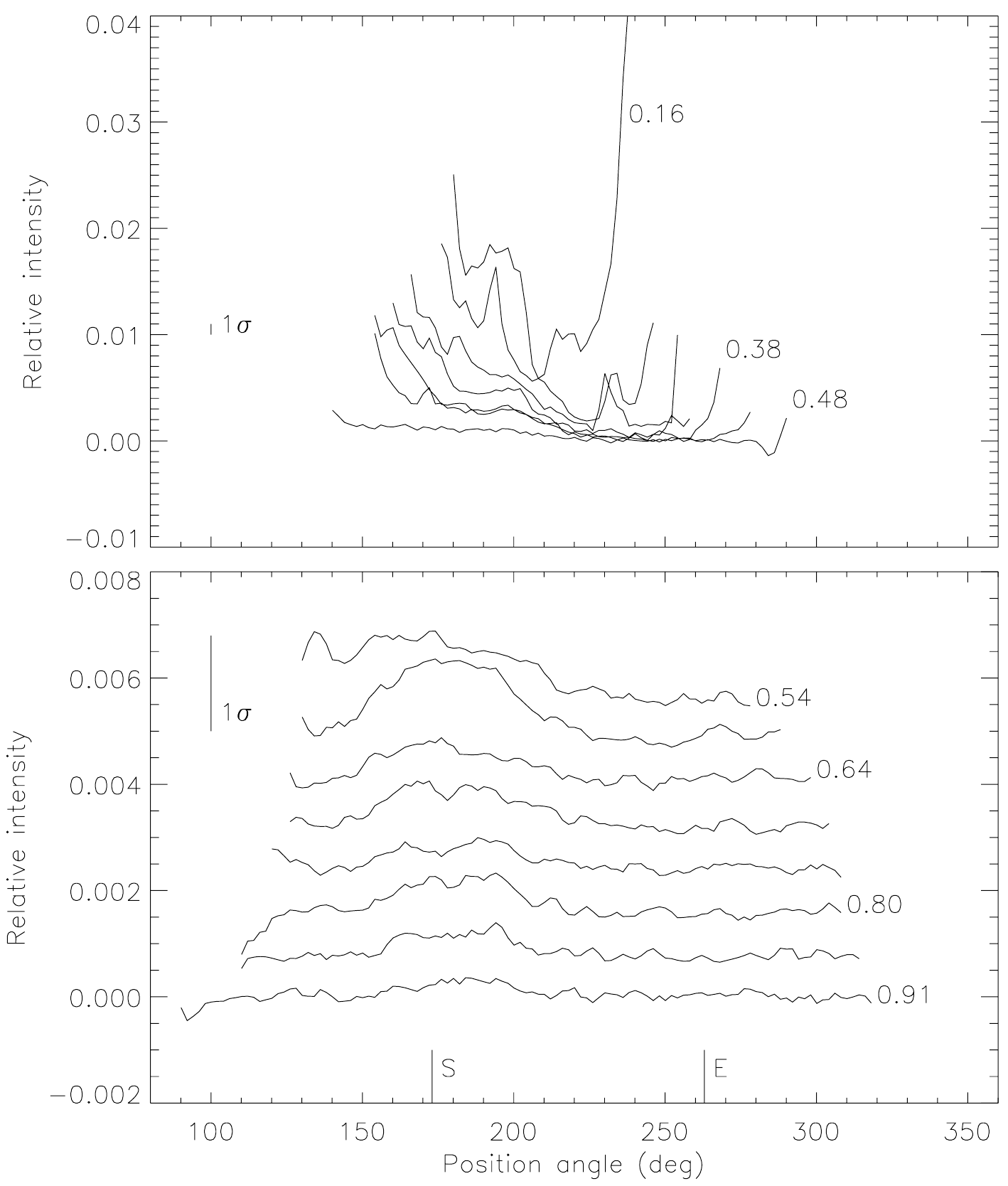}}
\caption{As in Fig.~\ref{F:DOT_Gband} in the CaII band at 396.8 nm.}
\label{F:DOT_CaIIH}
\end{figure}

For several reasons, DOT data constitute the most valuable dataset measured
for this work, thanks to the optimization of the instrument for solar observations \citep{Bettonvil03, Rutten04}. First of all, the image sampling (0.07 arcsec/pixel) provides a
comfortable scale for measurements, resulting in a Venus disk with a
radius of 415 pixels. Also, images in four narrow bands (usually employed 
for solar studies) are available, all having a FWHM of 1 nm or less (Table~\ref{T:obs}). 
Four cameras were working simultaneously in the four bands, taking every
minute an image burst of 100 frames at 6 frames/sec. 
For our measurements, all frames of each burst have been aligned and summed up
to obtain a single image each minute. Due to very strong turbulence the 
alignment process has been very critical, and was obtained by applying a
Sobel edge enhancement filter with a threshold and then using the Hough 
transform \citep{HoughTransform} to find the precise position of the disk of Venus in each image.

Fluxes were obtained for both the aureole and the background with the same method as with the previous data set. A step of 2 degrees along the Venus limb was used for measurements.
We show in Fig.~\ref{F:DOT_Gband} the photometric profile referred to the G
band. This result is very similar to both the blue and red continuum
images (Figures~\ref{F:DOT_BlueCont}-\ref{F:DOT_RedCont}). 
Only in the Ca~II~H band (Fig.~\ref{F:DOT_CaIIH}) some differences appear under
the form of additional peaks in the profile. Although a signature of 
high-contrast chromospheric structures in their refracted image cannot be completely ruled out, those features are most probably related to the
inner corona extending outward from the solar limb and contaminating the signal of the aureole, since its flux cannot be easily excluded from the measurement areas. 

However, as far as the main features are concerned, they are very similar in
all the four bands. For example, in all of them the shape of the profile soon
after the third contact (first curves in the upper panel,
Fig.~\ref{F:DOT_Gband}) has a similar shape, showing a bump at high negative latitudes.
This aspect is consistent with the observations obtained at the Themis telescope. 
This maximum of luminosity at later phases is located between 160 and 210
degrees of position angle, and appears to have a double peak.

Compared to the Themis data, the DOT curves show an overall lower signal. Their internal consistency and the difficulties encountered in measuring the Themis images affected by poor seeing suggest that the DOT measurements are probably much more reliable. Also, the low-resolution images by Comolli confirm the flux values obtained for the DOT. For these reasons, this data set is chosen for the detailed analysis in the following.

\section{Lightcurves and colors}
\label{S:color}
\begin{figure}[htbp]
\centerline{\includegraphics[width=0.5\textwidth]{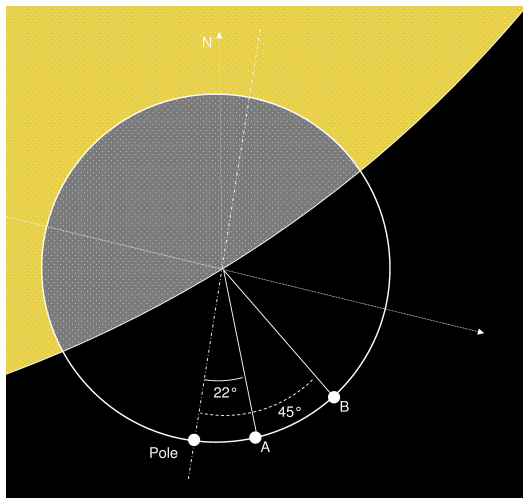}}
\caption{Geometry at $f=0.5$ during the final phases of the event. The large dots indicate the points that 
where chosen for studying the variation of the arc brightness in time. $A$ corresponds to the brightest portion,
and $B$ to the faintest.
}
\label{F:geom2}
\end{figure}

In order to study a possible wavelength dependence of the aureole and its variation in time, we used the DOT images for the egress phase.
The flux was measured on three arc segments centered on the pole, on the
brightest part of the aureole at position angle 195$^\circ$ (position A) and on
the faintest part at 218$^\circ$ (position B) as shown in the scheme
of Fig.~\ref{F:geom2}. Point B corresponds to a region at latitude 45$^\circ$S.

\begin{figure}[htbp]
\centerline{\includegraphics[width=0.55\textwidth]{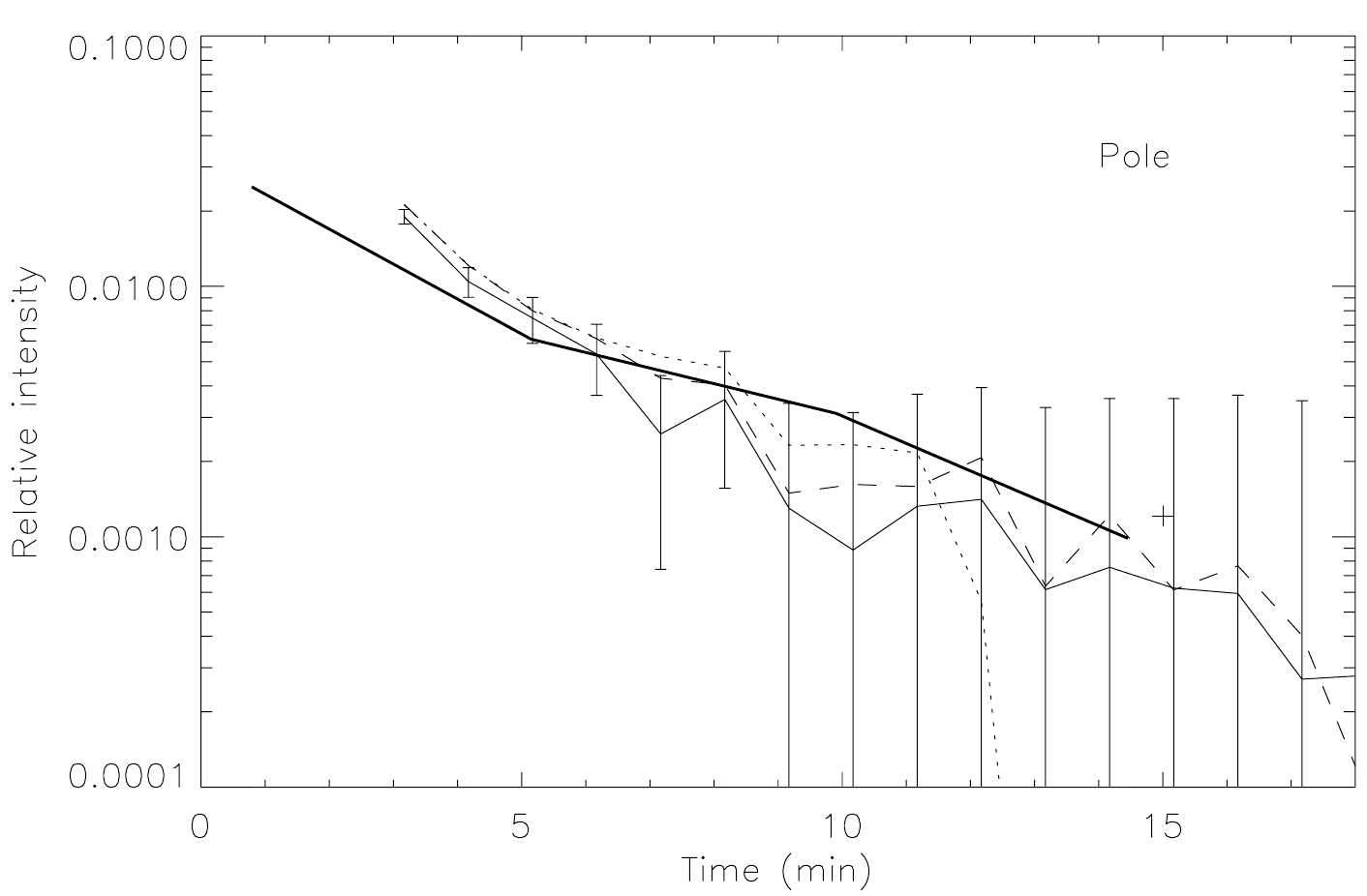}}
\centerline{\includegraphics[width=0.55\textwidth]{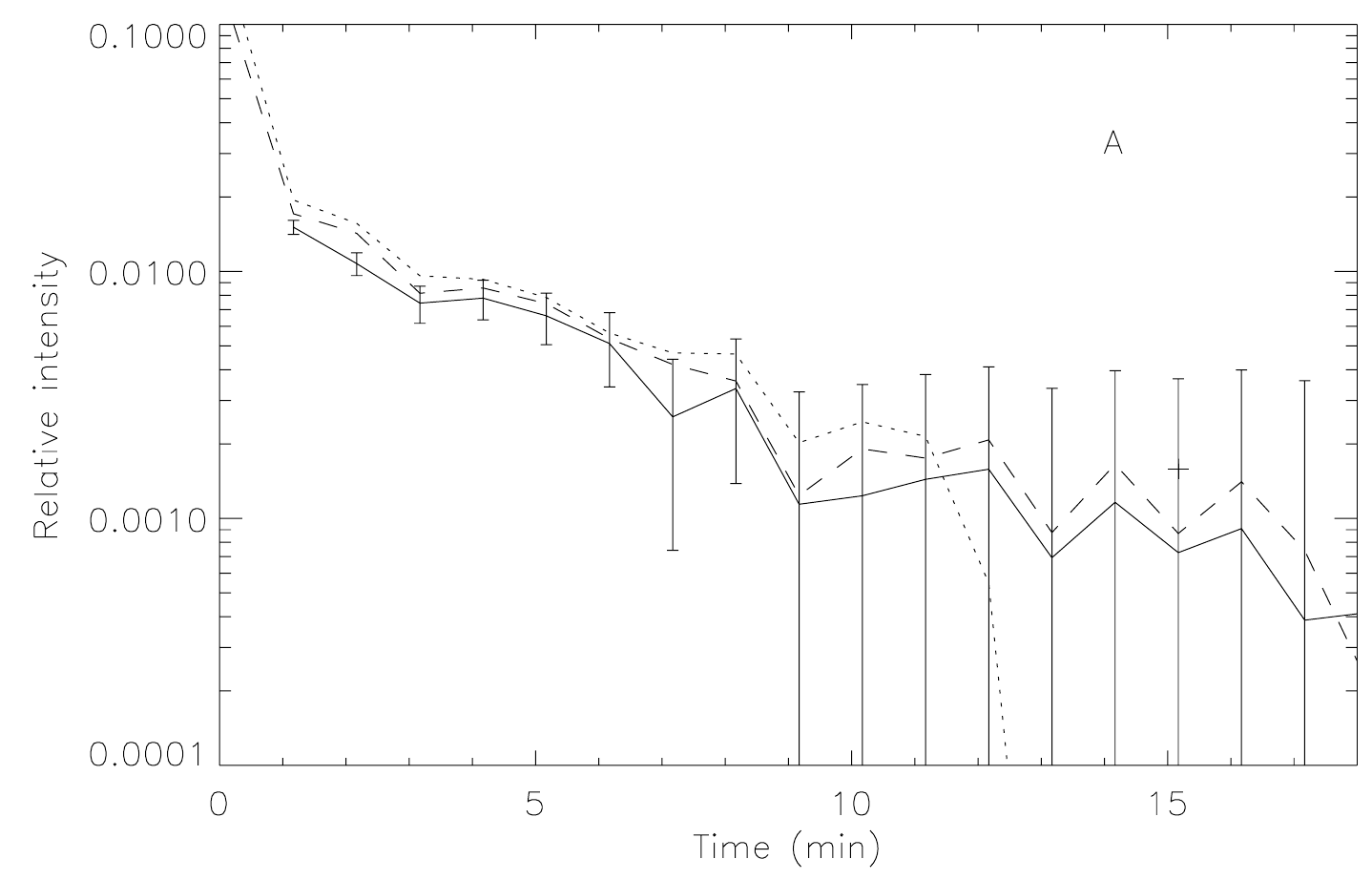}}
\centerline{\includegraphics[width=0.55\textwidth]{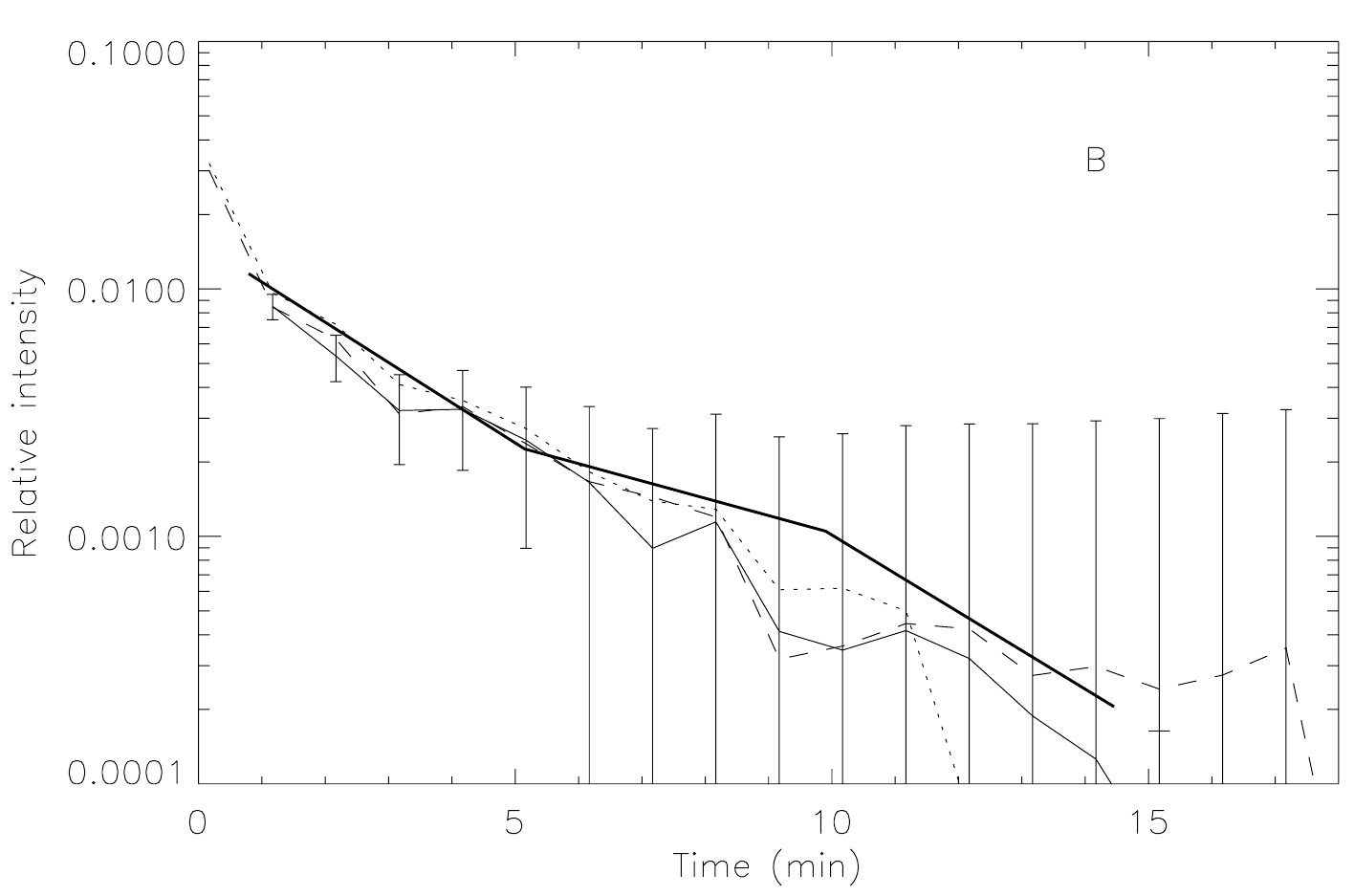}}
\caption{Relative intensity of the arc brightness in the DOT images, normalized as in the
  previous plots, as a function of time, where the origin of time corresponds to the third contact 
  (at 11:07:25 UT). The three panels refer to the Pole and to points A, B (in the order, from top to bottom). The G band, red continuum, and blue continuum are represented by the solid, dotted and dashed line respectively. The model fit is the thickest solid line in the top and bottom panel. At the B point, H=3.1 km and $\Delta$r=26 km provide the best fit, while at the pole the curve corresponds to H=4.8 km and $\Delta$r=38.5 km (see Sect. \ref{S:model} for details). The cross at t$\sim$15~min represents an isolated red measurement available. Error bars are very similar for all the three bands.
}
\label{F:time_multi}
\end{figure}

The measurements were made on arc lengths of 10 degrees, in the G band, blue
and red continuum. The results are shown in Fig.~\ref{F:time_multi}. 
On all the three plots, different wavelengths appear to follow a similar
trend. This is expected especially for the blue continuum and the G band, which are very close in wavelength. 

The fastest fall of flux in red continuum beyond t$\sim$12 minutes is a feature common to all the light curves, suggesting a certain chromaticity of the aureole. However, the related error bars are very large in this interval. At the same 
time, images around t~$\sim$~9-10~min are affected by a momentary seeing worsening, so even the color differences at these epochs are to be considered with caution. Also, the divergence of the blue and G fluxes beyond t~$\sim$~11~min is almost certainly spurious (due to the similar wavelength) and further underlines the difficulty of comparing very low fluxes.

In conclusion, we cannot firmly state that the images at our disposal reveal
a significant departure of the aureole color from the solar spectrum.

\section{Modeling the aureole}
\label{S:model}

\subsection{The refraction model}

The aureole observed during Venus transits can be explained by the refraction 
of solar rays through the planet outer atmospheric layers.
The rays that pass closer to the planet center are more deviated by refraction than those passing further out. 
The image of a given solar surface element is flattened
perpendicularly to Venus' limb by this differential deviation, while conserving the intensity of the rays, i.e.
the brightness of the surface element per unit surface. This holds as long as the atmosphere 
is transparent, i.e. above absorbing clouds or aerosol layers.

The refractive deviation of light is related to the physical structure of the planet atmosphere. 
The formalism of this deviation has been studied by Baum and Code (1953), for the cases
of stellar occultations by planets. Their approach assumes that 
$(i)$ the local density scale height $H$ of the atmosphere is constant and much smaller than the planet radius $R_v$, 
$(ii)$ the atmosphere is transparent, 
$(iii)$ and it has spherical symmetry.

This approach is still valid in our case, but has to be modified to account for the finite distance and size of the Sun. In the following we will consider that the physical 
properties of the transparent atmospheric layers vary smoothly with altitude, 
and remain about constant over a scale height.
Note that  the assumption $H \ll R_v$ is valid here, as $H$ is smaller than the atmospheric layer of Venus, in turn $\ll$ the planet radius.
Note also that the locally spherical symmetry is achieved for Venus' atmosphere. 
As to the transparency assumption, it must be dropped when the rays are going deep enough for
the atmosphere to become opaque.

\begin{figure}[htp]
\centerline{\includegraphics[width=0.95\textwidth]{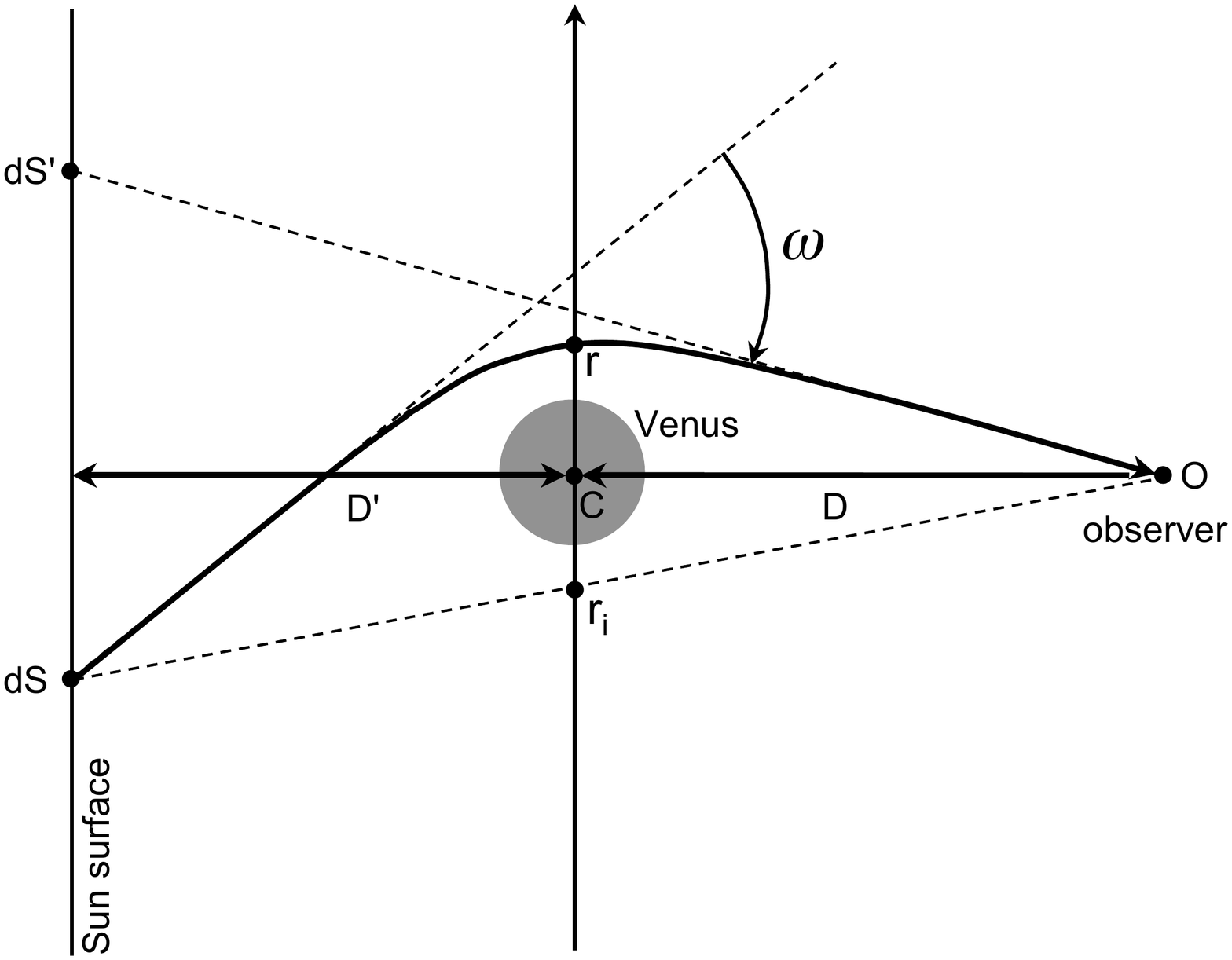}}
\caption{
Geometry of the refraction of solar rays by Venus' atmosphere, 
where $D'$ (resp. $D$) is the distance of Venus to the Sun (resp. Earth).  
A ray emitted from the solar surface at $dS$, is deviated by an angle $\omega$ 
(here negative) before reaching the observer $O$, who observes the images $dS'$
just above Venus' limb (the aureole).
The observer sees $dS$ projected at algebraic position $r_i$ in the plane going through 
Venus and perpendicular to the line of sight. The origin of $r_i$ is at Venus' center $C$ and
$r_i$ increases upward, see also Fig.~\ref{F:aureole}.
All sizes and angles have been greatly increases for better viewing. 
}
\label{F:refra}
\end{figure}

\begin{figure}[hbp]
\centerline{\includegraphics[width=0.95\textwidth]{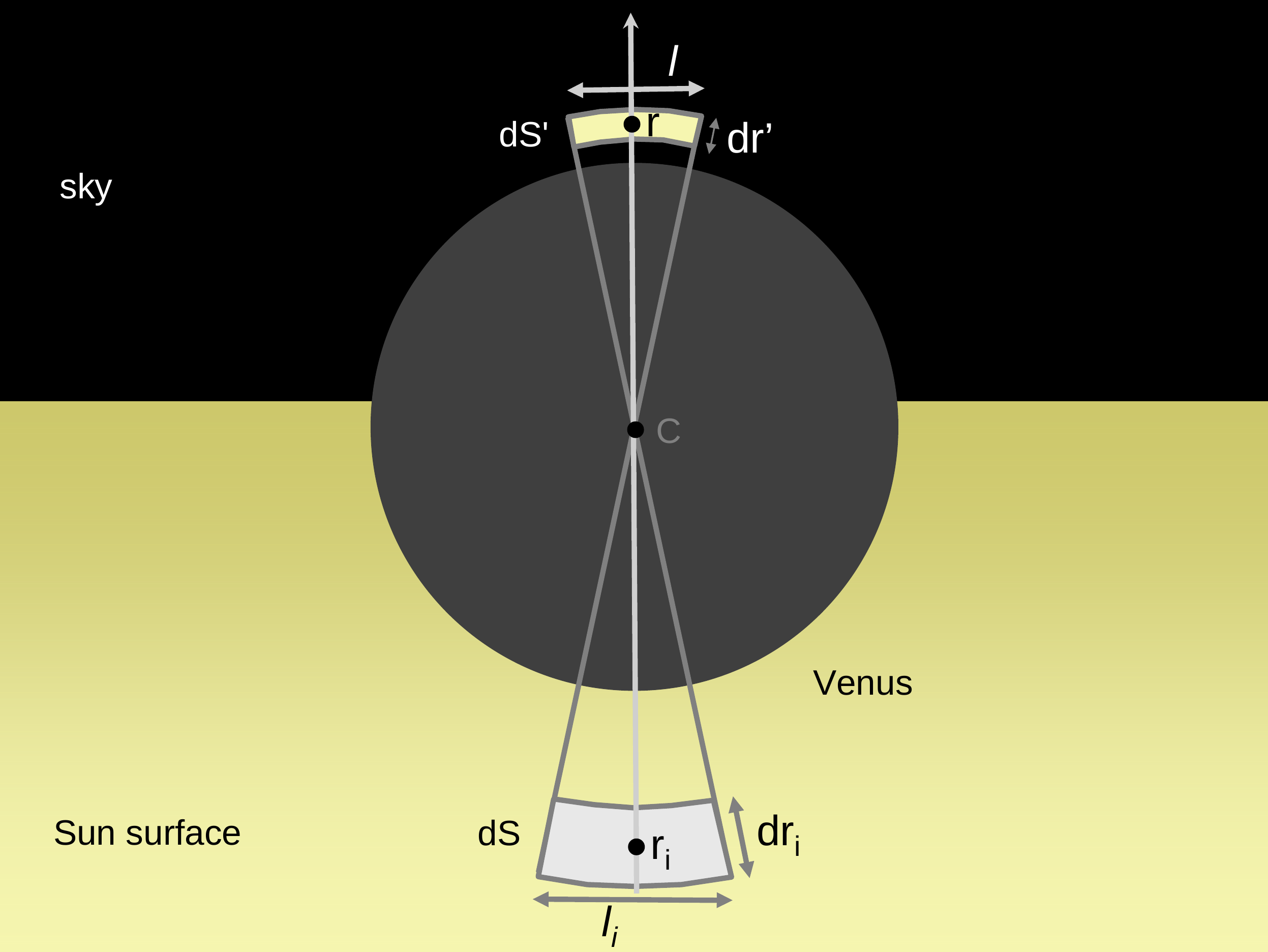}}
\caption{
Venus (dark gray disk) observed from Earth, partly against the solar disk and
partly against the sky (black background).
Each  solar surface element $dS$ surrounding $r_i$ has a  refracted image
$dS'$ of length $l$ and width $dr'$, caused by Venus atmospheric refraction. 
The image $dS'$ has the same surface brightness as $dS$ if the atmosphere is transparent.
See text for details.}
\label{F:aureole}
\end{figure}

We model the aureole brightness as follows:
 
(1)
We consider a surface element $dS$ on the Sun. A ray emitted by $dS$ will reach
the observer $O$ after being refracted by an angle $\omega$ in Venus's atmosphere (see Fig.~\ref{F:refra}). Note that we adopt here the convention $\omega \le 0$.
To each element $dS$ corresponds an image $dS'$ that will appear as
an aureole near Venus's limb (Fig.~\ref{F:aureole}).

During its travel to Earth, the ray passes at a closest distance of $r$ from the planet center.
Furthermore, $dS$ projects itself at a distance $r_i$ from Venus' center $C$, as seen from
$O$. Note that $r_i$ is an algebraic quantity which is negative if $dS$ and $dS'$ project themselves on opposite side relative to the planet center $C$,
and positive otherwise, see Figs.~\ref{F:refra}-\ref{F:aureole}.

(2)
The refraction angle $\omega$ is given by Baum and Code (1953):
\begin{equation}
\omega = -\nu(r) \sqrt{2\pi r/H},
\label{E:refrac}
\end{equation} 
where $\nu$ is the gas refractivity, decreasing exponentially with $r$.
This quantity is related to the gas number density $n$ by 
$\nu= K \cdot n$, 
where $K$ is  the specific refractivity. For  $CO_2$, 
we have\footnote{see http://www.kayelaby.npl.co.uk/general\_physics/2\_5/2\_5\_7.html}
$K= 1.67 \times 10^{-29}$ m$^3$ molecule$^{-1}$.

(3)
It can be shown that surface element $dS'$ is radially shrunk with respect to $dS$ by a factor
$\Phi= 1/\left[1 + D\cdot \left(\partial \omega/\partial r\right)\right]$, where $D$ is the distance
from Venus to Earth.
Note that since the atmosphere is assumed to have a constant density scale height,
$\partial \omega/\partial r = -\omega/H$, thus:
\begin{equation}
\Phi= \frac{1}{1- D\omega/H}
\label{E:phi}
\end{equation}

(4) It is convenient to take as a reference radius the closest approach distance $r_{1/2}$
corresponding to $\Phi=0.5$ (so so-called ``half-light radius" in the stellar occultation context).
Thus $\omega_{1/2} = -H/D$. Re-arranging the various 
equations given above, one finds $K \cdot n_{1/2} = \sqrt{H^3/\left(2\pi r_{1/2}D^2\right)}$,
where $n_{1/2}$ is the gas number density at $r_{1/2}$.
The latter equation permits to derive the numerical value of the half-light radius, once an
atmospheric model of the planet is given, that is, once the density profile $n(r)$ is specified.

(5)
From Eq.~\ref{E:refrac}, we derive:
\begin{equation}
\omega = -\frac{H}{D} \cdot e^{-\left(r-r_{1/2}\right)/H}
\label{E:omega}
\end{equation}
Simple geometrical consideration show that we also have:
$\omega = -k\left(r-r_i\right)/D$, where $k= 1 + D/D'$ ($D'$ being the distance of Venus to the Sun).
Combining this equation with Eqs.~\ref{E:phi}-\ref{E:omega}, we obtain:

\begin{equation}
\frac{1}{k}\left(\frac{1}{\Phi} - 1\right) + \log\left(\frac{1}{\Phi} - 1\right)= \frac{r_{1/2}-r_i}{H}
\label{E:BC}
\end{equation}

This is the Baum and Code formula (apart for the correcting factor $k$, 
which is equal to unity in the original formula  since $D'= +\infty$ for stellar occultations).

Thus, for each value of $r_i$, we can calculate the corresponding value $\Phi(r_i)$,
using a classical Newton iterative numerical scheme.

For each value of $r_i$, we can also calculate the corresponding closest approach radius $r$
by combining Eqs.~\ref{E:phi} and \ref{E:omega}: 
\begin{equation}
r= r_{1/2} - H \cdot \log\left(\frac{1}{\Phi} - 1\right)
\end{equation}

To calculate the flux $dF$ received from an aureole element of surface 
$dS'$ with length $l$  and width $dr'$ (Fig.~\ref{F:aureole}), 
it is enough to note that the surface brightness of $dS$ and $dS'$ are the same
if the atmosphere is transparent, i.e. that
$dF= S_\odot(r_i) l dr'$, where $S_\odot(r_i)$ is the flux 
received from a unit surface on the Sun at $r_i$ (taking into account the limb-darkening effect).  
Thus, we can re-write this equation as $dF= S_\odot(r_i) \Phi(r_i) l_i dr_i$. 

The aureole is not radially resolved, so we have only access to the flux integrated
along $r_i$, i.e. to:

\begin{equation}
F= \int_{r_{i,\rm{min}}}^{r_{i,\rm{max}}} S_\odot(r_i)  \Phi(r_i) \cdot l_i \cdot dr_i
\end{equation}

The lower bound of the integral, $r_{i,\rm{min}}$, corresponds
to the value of $r$ corresponding to an opaque cloud or aerosol layer at altitude r$_{cut}$.
The upper bound of the integral, $r_{i,\rm{max}}$, corresponds
the solar limb, beyond which no more photons are emitted.

By applying this model, in the following we will determine the scale height $H$ and the half occultation radius relative to slanted opacity $\tau\sim$1 ($\Delta r = r_{1/2} - r_{cut}$) best reproducing the observations. In general, different portions of the arc can yield different values of these parameters, thus providing a useful insight of variations in the physical properties of the Cytherean atmosphere as a function of latitude.

\section{Derivation of the physical parameters}
\label{S:application}

We used the equation (\ref{E:BC}) for modeling the refraction in the atmosphere of Venus and the flux in the observed aureole. Essentially, the mathematical model depends upon two parameters: the altitude of the half-occultation level relative to an underlying totally opaque layer ($\Delta$r) and the optical scale height of the atmosphere H, second free parameter of Eq.~\ref{E:BC}.
\begin{figure}[htp]
\centerline{\includegraphics[width=0.95\textwidth]{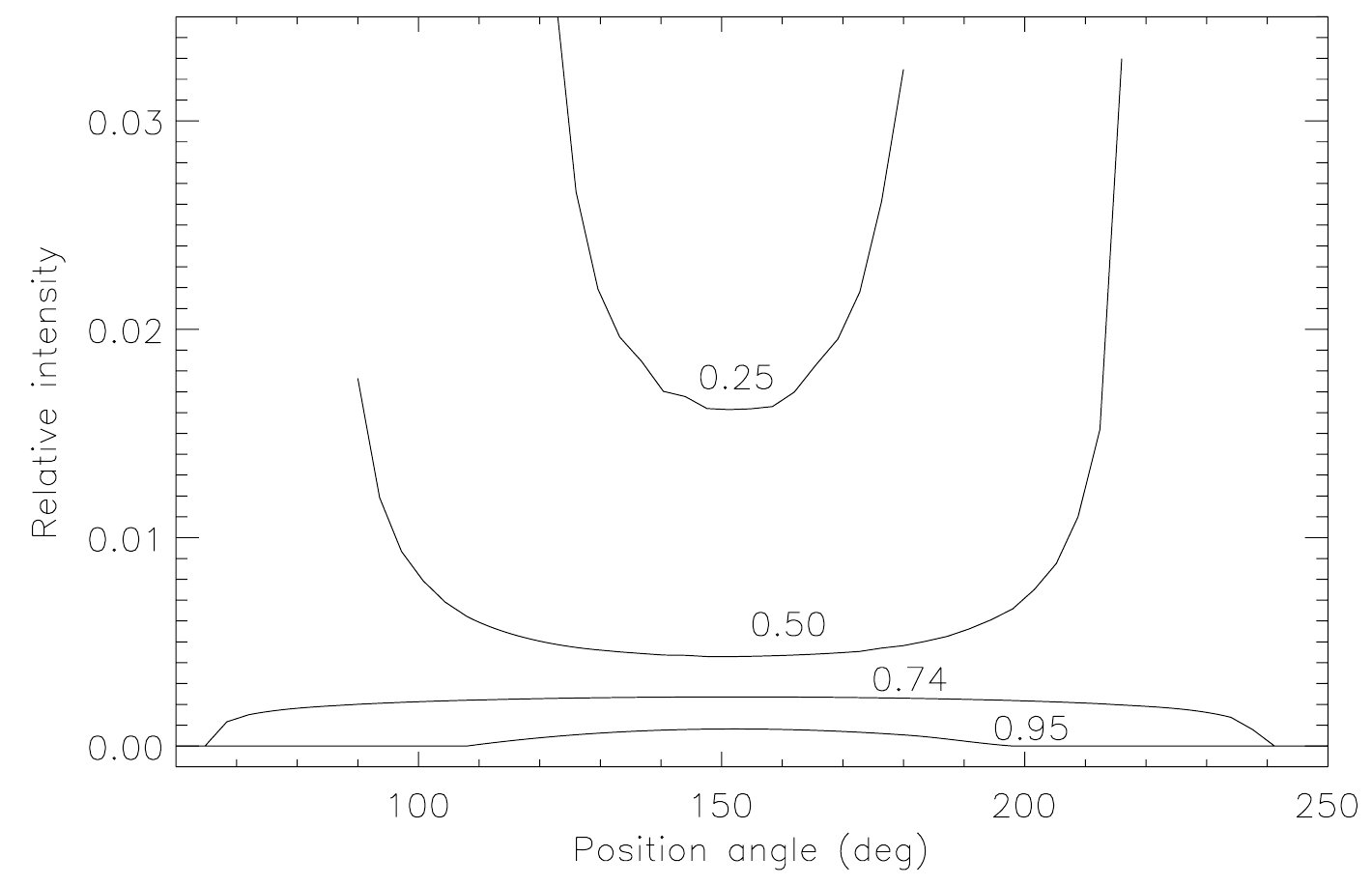}}
\caption{Relative intensity of the arc brightness for H=4.8 km, 
$\Delta$r$\sim$38.5 km as derived from the model. Labels correspond to the $f$ fraction of disk overlap.}
\label{F:model}
\end{figure}

Since our brightness measurements are referred to different epochs, the model must fit at the same time not only single-epoch profiles, but also their evolution. Although free parameters of Eq.~\ref{E:BC}, r$_{1/2}$ and H are quantities that are not physically independent since they are both related to local physical structure of the mesosphere of Venus. We will thus check afterwards the physical consistency of our results.

The source of refracted rays (the solar photosphere) is computed as a smooth function of the radial photometric profile of the Sun (given by~\citealt{hestro98}), whose parameters are determined by a fit to the profile measured on the solar photosphere imaged close to the Cytherean disk.

Fig.~\ref{F:model} shows a typical outcome of the model for different phases $f$ of Venus ingress or egress, considering the atmospheric properties to be constant all along the planet limb. When Venus appears largely overlapped to the solar disk, the refracted flux is dominated by the arc extremes, where it approaches the solar limb. This is a consequence of a very small bending of the light path occurring high in the atmospheric layer involved. When f~$>$~0.5, the regions of the photosphere contributing to the arc brightness are all farther away than the Venus disk center -- so the larger deviation associated to a deeper pass of the solar light inside the atmosphere is needed. In such situation, the arc extremes are fainter than the center, since their brightness sources -- being on the opposite side of the planet center -- are on the strongly darkened photospheric limb. For higher values of $f$ the signal rapidly fades. The cut-off at zero flux is a feature of the model due to the opaque layer, blocking the refracted rays that cross the atmosphere below a given altitude.

\begin{figure}[htp]
\centerline{\includegraphics[width=0.8\textwidth]{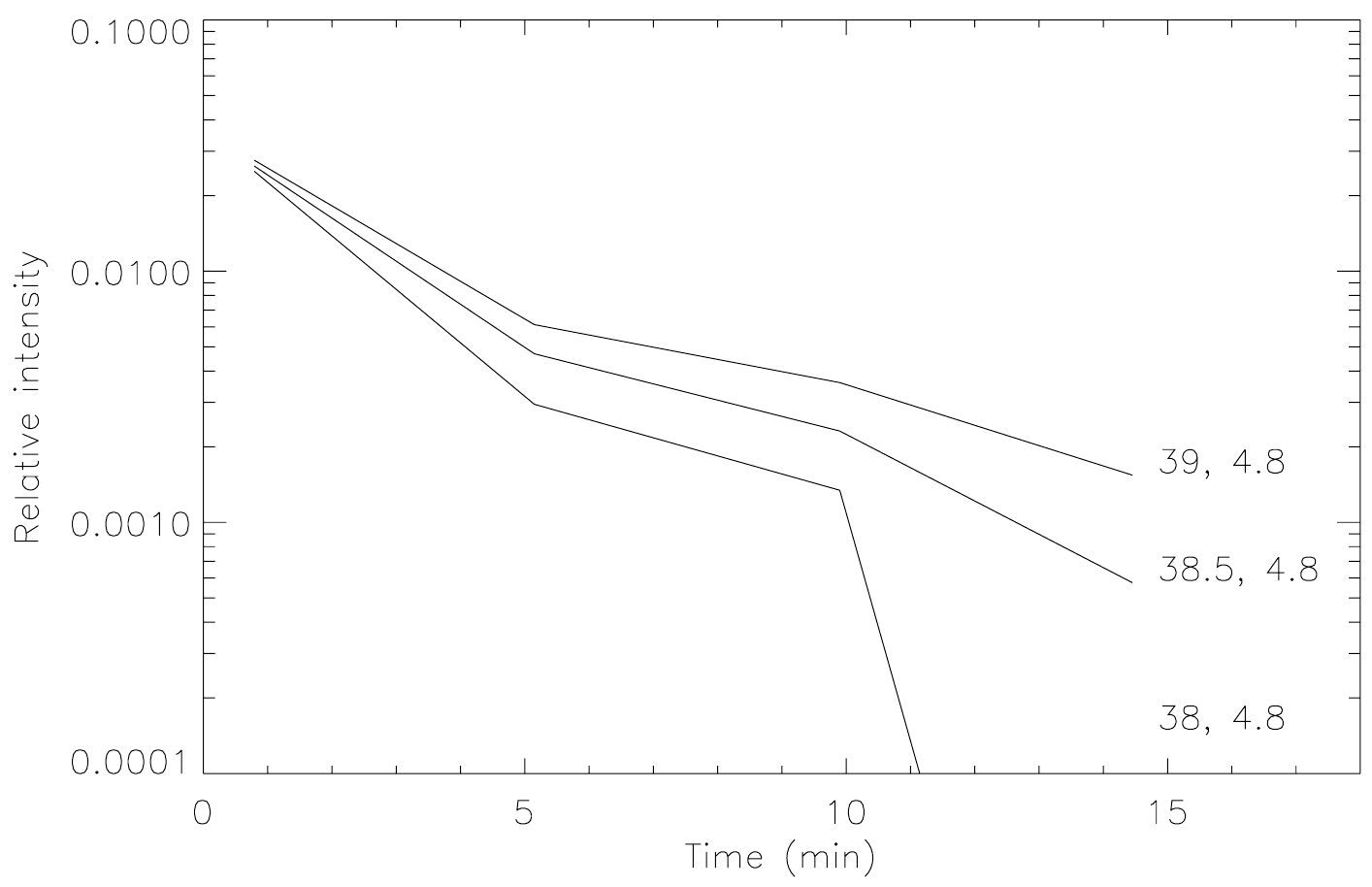}}
\centerline{\includegraphics[width=0.8\textwidth]{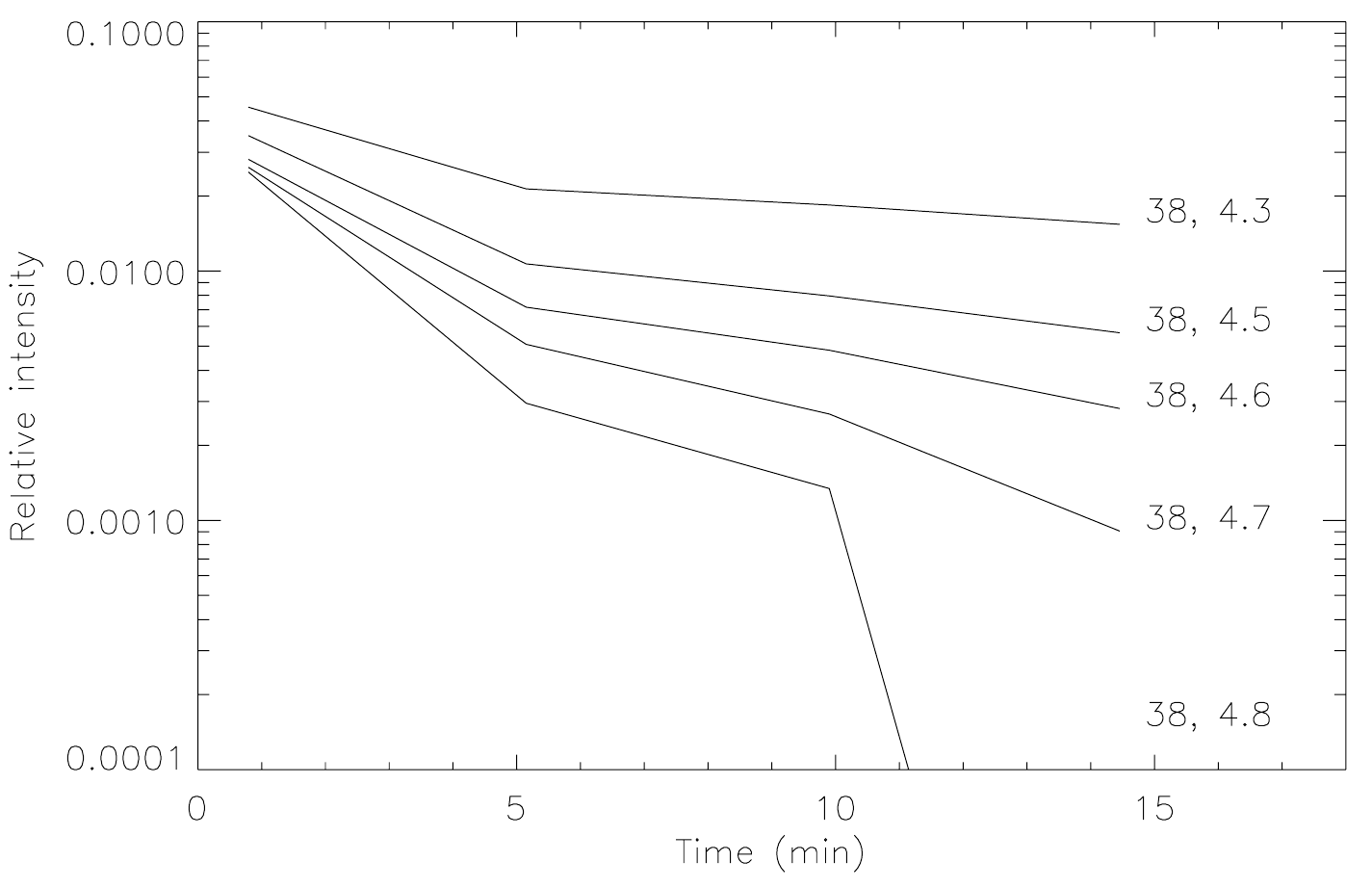}}
\caption{Effects of the variation of the physical parameters controlling the aureole flux. In the top panel, H$_{eq}$=4.8 km is kept constant, while the three simulated light curves correspond to different values of $\Delta$r$_{eq}$ (indicated by the labels). On the other hand, in the bottom panel $\Delta$r$_{eq}$=38 km is constant and H is varied.}
\label{F:changing}
\end{figure}

Diminishing $H$ determines a stronger refraction that increases the amount of light coming from regions that are farther away from Venus on the solar photosphere. The slope of the flux decrease that is observed when Venus exits the photosphere (i.e. when $f$ is increasing) will also be less steep (Fig. \ref{F:changing}). 

Another feature of the model outcome is the symmetry relative to the axis joining the center of the disks of Venus and the Sun. It is thus clear that the asymmetries and features visible in the measurements must be related to latitude variations of the physical properties in the refracting atmospheric layers.

We searched for the values of $\Delta$r and H that better fit the time variations reproduced in Fig.\ref{F:time_multi}, separately for the Pole (or, equivalently, the point A) and the point B. The procedure we adopted started with a search for the value of $\Delta$r, which mainly affects the amount of aureole flux at small $f$  (i.e. when Venus is nearly completely on the solar disk). Then, H is varied in order to obtain the right slope for the flux variation in time. The procedure is repeated until a convergence to the observed data is obtained. 

The central portion of the arc, for which the longest time series is available, is best fitted by H$_{eq}$=3.1 km and $\Delta$r$_{eq}$=26 km (Fig.~\ref{F:time_multi}). This region corresponds to intermediate latitudes ($\sim$45$^\circ$ - point B). 

We then fitted the Pole area, obtaining H$_{pole}$=4.8~km and $\Delta$r$_{pole}$=38.5~km. This determination appears less satisfactory at low $f$ values (beginning of egress). If a better fitting of this part of the curve is forced (in particular by increasing $\Delta$r) the central plateau of the curve becomes more extended, and the fit is worsened in the rest of the domain. We thus preferred to privilege the overall better fit. This choice appears to be reasonable when compared to the expected altitude difference of cloud top and upper haze between the equator and poles (see next section).

In fact, if refraction occurs at similar absolute altitudes in the atmosphere, variations of $\Delta$r should only correspond to changes in the altitude of the opaque layer, i.e. a slanted opacity (optical thickness) $\tau\sim$1. Our best fit values indicate that the higher flux received at the polar latitudes on Venus can be related to a significantly lower aerosol extinction towards the poles, thus allowing more refracted rays to reach the observer. This contribution is blocked at the equatorial latitudes by the higher clouds.

An indication of the formal accuracy of the determination of $\Delta$r and H can be obtained by studying the fit sensitivity to changes in the parameters. We have verified that a change of $\sim$3~km in $\Delta$r and $\sim$100~m in H are sufficient to displace the fitting curve of an amount exceeding the extension of the error bars all along the lightcurves. The rapid variations in the aureole behavior following changes in the parameters are illustrated in Fig.\ref{F:changing}. In particular, our results appear extremely sensitive on H.
As a further consistency check, we can estimate $\Delta$r from Eq.~\ref{E:omega} by using a value of H$\sim$4~km and assuming the maximum deviation of light rays (corresponding to a highest $f$ at which the signal from the aureole is detected) to be of the order of $\omega$=40~arcsec. In this case $\Delta$r$\sim$35~km, very close to the results given above obtained by a full model fit to the observed data. 

Eventually we can estimate the visibility of the aureole in terms of visual magnitude.  We use the simple law for the profile of the solar limb darkening (normalized to the Sun's center brightness) by \citet{hestro98}:
\begin{equation}
I(\mu)=1-u\left(1-\mu^\alpha\right)
\label{E:eqsun}
\end{equation}
in which $\mu=\left(1-r^2\right)^{0.5}$. Here, $r$ represents the distance from the center of the Sun, normalized to the solar radius. At a wavelength in the V-band range ($\lambda$=579.9~nm) this law fits the observed profiles when u=0.85 and $\alpha$=0.8. In the case of our reference area the value of r is r$_v$=0.969, so I$_v$=0.428. In this region so close to the solar limb the profile approximation can be less accurate, but it should not be erroneous by more than 2$\%$ \citep{hestro98}. 

From the computed I$_v$ and using the magnitude of the Sun (V = -26.71) we obtain a surface magnitude per arcsec$^2$ for the reference element m$_v=-10.87$. By observing the plots, we can assume that the aureole has been easily seen and imaged when its brightness was $\sim$10$^{-2}$ times the surface brightness of the reference area (1 arcsec$^2$ at 1 Venus radius from the solar limb), i.e. 5 magnitudes fainter. We can thus consider that the typical magnitude of an arc of aureole 1~arcsec in length was -5.9.

One should note, however, that ample variations around this value are present, both along the arc and over time. 

\section{Comparison to Venus Express observations}
\label{S:VeX}
Venus' mesosphere, which extends typically from 70 to 110 km, appears to be a transition region between the troposphere (0--70 km), dominated by the $\sim$4 Earth days retrograde zonal superrotation, and the thermosphere (110--~250 km) in which solar EUV and diurnal temperature contrasts drive a predominantly subsolar-to-antisolar (SS--AS) circulation. Monitoring of the mesopheric thermal structure has revealed that this region is characterized by a strong temporal variability, whose origin remains poorly understood \citep{Bougher97, Bougher06, Lellouch97, Lellouch08, Sandor05, Gurwell07, Widemann08}. Upward from cloud top near 70 km, zonal winds decrease with height while thermospheric SSÜAS winds increase. The region further serves as the primary photochemical region of the Venus atmosphere and heterogeneous chemistry revealed by haze and cloud particles production. Although present in small amounts, trace gases are involved in important complex chemical cycles. Photochemical reactions between CO$_{2}$, SO$_{2}$, H$_{2}$O and chlorine compounds lead to the formation of sulfuric acid, which is the main component of the cloud and upper haze, associated with low H$_{2}$SO$_{4}$ vapor pressure \citep{Zhang10}. 

\subsection{Cloud and haze structure}

At the limit between the troposphere and the mesosphere, the upper cloud structure reveals distinct dynamical structure in the latitudinal direction, with convective, wave-dominated zonal flow in the lower latitude range near Venus's equator, and a significant transition to less convective banded flow between 45$^\circ$ up to about ±70$^\circ$--80$^\circ$. The temperature field forms a bulge of cold air at 60$^\circ$--80$^\circ$ latitude called the "cold collar region", which vertically extends up to 75 km. The torus-like structure encloses a vast vortex several thousand kilometers across, with a slower rotation period ($\sim$2.5 to $\sim$2.8 Earth days) and significantly depressed cloud top altitude, at about 62--65~km instead of 70--74~km closer to the equator~\citep{Piccioni07}. The structure of the cloud tops is especially poorly investigated since it falls between the altitude ranges sounded by solar/ stellar occultations and that studied by descent probes. Recent analysis has been performed on board Venus-Express~\citep{Svedhem07} based on depth of CO$_{2}$ bands at 1.6 $\mu$m measured by VIRTIS (Ignatiev et al. 2009), and CO+CO$_{2}$ gaseous absorption in the 4.5-5 $\mu$m  range using VIRTIS and VeRa temperature profiles (Lee et al. 2010) while~\citet{Luz11} brought the first extensive characterization of the vortex dynamics and its precession motion. Although the absolute cloud top altitude poleward and equatorward of the polar collar differs when applying the two modeling techniques  (63--69 km / 74 $\pm $ 1 km in Ignatiev et al., 2009 ; 62-64 km / 66 km in Lee et al., 2010), their relative difference is comparable (1-2 scale heights). The latitudinal extent of this polar depression matches well the brightest portion of the aureole observed southward of 65$^\circ$--70$^\circ$ near the South polar limb in DOT telescope data (see Fig.8a of Ignatiev et al. 2009). Interestingly, similar agreement can be traced back to Russell's drawings of 1874 (See Link, 1969 and Fig. 1 of Pasachoff et al., 2011).

\subsection{Haze aerosols}

The Venus upper haze (70--90 km) was first evidenced by measurements from Pioneer Venus orbiter limb scans at 365 and 690 nm at northern midlatitude. It is mainly composed of submicron sulfuric acid (H$_{2}$SO$_{4}$) aerosol particles with typical radii from 0.1 to 0.3 $\mu$m (Lane and Opstbaum, 1983 ; Sato et al., 1996).~\citet{Wilquet09} also present the first evidence for a bimodal particles distribution, similar to the two modes in the upper clouds, at the latitudes probed during the observation of solar and stellar occultations by ESA's Venus Express, with typical radii of Òmode-2Ó particles between $\sim$0.4 and 1$\mu$m. These measurements were performed close to the polar regions. From the  results of~\citet{Wilquet09} the absorption at different wavelengths as a function of the altitude can be deduced. In general, the absorption levels are found to be $6\pm1$~km higher in the visible domain (SPICAV-IR at 757~nm) than at 3~$\mu$m.

As an integrated aerosol optical depth $\sim$1 is measured a few density scale heights above the cloud tops, we note that the slanted geometry of the aureole must reach half occultation level $r_{1/2}$ well above the upper haze. Quantitatively, the altitude of the aureole's half--occultation level in the polar region will be found by adding the value of $\Delta$r$_{pole}$=38.5~km to the altitude where aerosols slanted opacity $\tau\sim$1.
Recent VEx/SOIR results \citep[][personal communication]{Wilquet11b} place that altitude at $73\pm2$ km in the $3\mu$m band, to be further increased by $6\pm1$ km to retrive its value in the visible domain, as above. The final sum thus yields r$_{1/2}\sim 117.5\pm 4$~km at the pole. Most recent VEx/SOIR results integrated over four years of observations (Wilquet et al., 2011) further indicate an altitude of longitudinally averaged, integrated aerosol optical depth $\sim$1 increasing toward the equator, with an altitude of $81\pm 2$~km for latitudes between 35$^\circ$S and 55$^\circ$S. Therefore, as $\Delta$r$_{eq} = $r$_{1/2} - $r$_{cut}$ is significantly smaller than $\Delta$r$_{pole}$ the resulting value of r$_{1/2}$ at mid-latitude differ from our calculation in the polar region by about two scale heights (Table 2). We can thus consider that our results only show marginal latitudinal variation of the altitude for the half-occultation level along the terminator, although in a region of important temperature variation which need to be independently assessed.
The same SPICAV/SOIR results show that, under reasonable assumptions, the refraction index at visible wavelengths is fairly constant, thus confirming that the process is not a source of relevant chromatic effects in the aureole.

\subsection{Temperature structure}

Only scarce measurements of vertical temperature profiles have been performed above 100 km altitude, where temperature fields, especially in the polar region, as well as their time variability, are still debated (see, e.g., Vandaele et al. 2008, Clancy et al. 2008, 2011; Piccialli et al. 2008, 2011). Inverted equator-to-pole temperature gradient on isobaric surfaces above 75 km, were first reported by NASA's Pioneer Venus Infrared Radiometer and radio occultation measurements (Taylor et al., 1983; Newman et al., 1984), Venera-15 Fourier Spectrometry data (Zasova et al., 2007) and more recently ESA's Venus Express, e.g. Grassi et al. (2008) ; Piccialli et al. (2008) using VIRTIS -M observations ; Tellmann et al. (2009) based  on VeRa radio occultations on board VEx. The collar region also divides the atmosphere vertically. Below the collar, the atmosphere cools with increasing latitude. Above, the temperature gradient is reversed, as diabatic heating or dynamically controlled processes could be responsible for the observed structures (Schubert et al., 1980, Tellmann et al., 2009). At the altitude of the collar and closer to the poles, the atmosphere is almost isothermal and also much warmer than the surrounding areas. Vertical temperature profiles are fairly constant with altitude between around 50 and 100 km, above which they begin to rise sharply in the mesosphere, see e.g. Fig. 2 in~\citet{Mueller06}. 

Retrieving the temperature using the density scale height should be made with caution. The transition between stable and adiabatic regions is fairly smooth at equatorial latitudes, but abrupt at middle and polar latitudes~\citep{Tellmann09}. A significant vertical temperature gradient is a potential issue when trying to compare the density scale height to the temperature scale height (see the discussion related to Titan's mesosphere refractivity in~\citealp{Sicardy06}). However, we consider the mid-occultation level probed by the arc in the $\sim$115 km region (see Table 2) as isothermal. This is an acceptable hypothesis since the diurnal average of dT/dz on the background atmosphere (Hedin et al., 1983) is negligible, although significant local-time and daily variability has been retrieved (Clancy et al. 2008, 2011; Sonnabend et al. 2008, 2011). In particular, the elevated value of T$_{iso}$ = 212K at 120.5$\pm$4 km at 68$^\circ$S is in agreement with high kinetic temperatures derived from heterodyne mid-infrared spectrum of non-LTE emission at terminator, see in particular Table 1 of Sonnabend et al. 2011. This result suggests a warmer or rapidly variable temperature condition in the altitude range.

\begin{table}[t]
\caption{Results of modelling and retrieved half-light altitude r$_{1/2}$, scale height H (in km) and temperature along the Lomonosov's arc, assuming a locally isothermal atmosphere.  Explanations of symbols in the text. T$_{\rm{iso}}$ = m(z) g(z) H/k. T$_{\rm{mod}}$~(K) from Hedin et al. 1983. \label{T:res}}
\begin{tabular}[t]{ll|cc|cccc}
\hline
Point & Lat. & $\Delta$r &  H  & r$_{cut}$ & r$_{1/2}$ & $T_{\rm{iso}}$~(K) & $T_{\rm{mod}}$~(K)\\
\hline
A & 68$^\circ$S & 38.5$\pm$3 & 4.8$\pm$0.1 & 76$\pm$2 & 120.5$\pm$4 & 212 & 138-190\\
B & 45$^\circ$S & 26.0$\pm$3 & 3.1$\pm$0.1 & 81$\pm$2 & 113.0$\pm$4 & 137 & 143-185 \\
\hline
\end{tabular}
\end{table}

\subsection{Physical interpretation}

In order to check the consistency of our results with the known physics of Venus atmosphere, we can compute the expected value of the density scale height H at the altitude of the refracting layer. The temperature scale height is defined as $H = kT (mg)^{-1}$, where k is the Boltzmann's constant, while T, m and g respectively represent temperature, molecular mass and gravitational acceleration at the altitude considered. As discussed above, we consider a locally isothermal atmosphere. For our computation we derive the value of g as a function of altitude. We derived an altitude for the opaque layer from SOIR aerosol transmittance measurements near 3 $\mu$m (Wilquet et al., 2011 ; Wilquet, personal communication), and added $6\pm1$~km for transferring to the visible domain. We then adopt the model by \citet{Hedin83} describing the atmospheric structure on Venus at noon and midnight, around the equator. We consider that an appropriate temperature estimation can be obtained by averaging the two profiles. Differences may be present, since we observe directly at the terminator, but the temperature data by SOIR obtained at a similar geometry \citep{Mahieux10} do not show significant discrepancies relative to the model. CO$_{2}$ is the dominant gas on the dayside below 160 km and on the nightside below 140 km, replaced at higher altitudes by O. Mean molecular mass remains close to 44 a.m.u. below 130 km, where CO$_{2}$  dominates, and decreases towards higher altitudes~\citep{Hedin83,Mueller06}. Although day--night differences in composition are substantial, we also may consider longitudinally averaged values at the altitude considered. Concerning $m$, since the layer is in the heterosphere, a decrease due to fractionation has to be taken into account, due to the decrease in CO$_2$ relative to other components, in particular atomic oxygen O, molecular nitrogen N$_{2}$ and He. As a result, at 120~km we use an average molecular mass of 42.9, and a day-night longitudinal average T=164 K, obtaining H=3.7~km. An equal value is obtained at 115 km. Both are in remarkable agreement with the average one obtained from the aureole measurements, H=3.95~km (see Table 2). Finally, we noted that our results only show marginal latitudinal variation of 1-2 scale heights for the altitude for the half--occultation level along the terminator, near 117 km. 

\section{Conclusions} 

Our results represent the first successful model of the aureole of Venus, observed during solar transits. We are able to reproduce the main features of the lower mesosphere observational constraints : longitudinally averaged thermal structure near 117 km, slanted opacity of aerosols and their meridional variation, with consistent physical parameters. 

The results obtained by this first set of measurements are encouraging and suggest that more accurate planning will produce more precise data. Improvements are also possible concerning the refraction model, which could be completed by a more realistic, gradual transition from transparency to absorption at the cloud deck level based on most recent results on the latitudinal distribution of upper haze aerosols. Also, the contribution of scattered light in addition to refracted light could be included. In fact, the absorption curves in \citet{Wilquet09} Fig. 3 are wavelength dependent, i.e. the opaque layer altitude in our model can be function of the color. As a result the aureole should not be perfectly ''grey'', suggesting that future observations capable of accurate multicolor photometry should investigate this possibility.Even a grey absorption would cause the solar flux in the aureole to look dimmer than without aerosols, which would be wrongly interpreted as a drop in scale height, i.e. in local temperature at tangent point altitude, in the inversion code.

In the two hemispheres, the cold collar temperature structure and the polar regions are very similar~\citep{Tellmann09}, so the latitudinal variations obtained for the 2004 transit around the South polar region might tentatively apply to the North polar data we expect to acquire in June 2012. Recent interpretations also attest a general N-S symmetry in the latitudinal variation of aerosols extinction \citep{Wilquet11b}. Given the difference in brightness between the arc and the solar photosphere, accurate measurements of this phenomena are challenging. The degree of turbulence, the magnification, the amount of light scattering in the optics are all elements that contribute in determining the effective visibility of the aureole in a given instrument, and the accuracy of the measurements. The transit of June 2012 will be our last opportunity for observing the aureole using this spatially resolved technique, until the next pair of transits of Venus, which will be in the ascending node, on 11 December 2117 and 8 December 2125. 

\section*{Acknowledgments}

We thank the referees for useful comments on the manuscript, M. Frassati and Unione Astrofili Italiani for the use of the drawing reproduced in Fig.~\ref{F:frassati}. We also thank Val\'erie Wilquet, Ann-Carine Vandaele and Arnaud Mahieux of the Belgium Institute for Space Aeronomy for sharing their work in progress on the longitudinally averaged optical extinction of mesospheric aerosols on Venus. R. Hammerschlag (Dutch Open Telescope, La Palma) provided valuable help and comments.

\section*{References}
\bibliographystyle{elsarticle-harv}







\end{document}